\documentclass[a4paper,11pt]{article}
\pdfoutput=1
\usepackage{jheppub}
\usepackage[T1]{fontenc}
\usepackage[latin1]{inputenc}
\usepackage[english]{babel}
\usepackage{amsthm}
\usepackage{empheq}
\usepackage{mathdots}

\DeclareMathOperator{\diag}{diag}
\DeclareMathOperator{\adj}{ad}
\DeclareMathOperator{\rank}{rank}
\DeclareMathOperator{\ord}{ord}
\usepackage{dsfont}
\usepackage{pifont}

\usepackage{ulem}
\hyphenchar\font=\string"7F
\usepackage{bbold}
\usepackage{colortbl}
\definecolor{fillcolor}{RGB}{196,196,196}
\usepackage{tikz}
\usetikzlibrary{calc,matrix,scopes,arrows}
\tikzset{>=stealth'}

\title{\boldmath Flux Formulation of DFT on Group Manifolds\\and
Generalized Scherk-Schwarz Compactifications}
\preprint{LMU-ASC 60/15\\MPP-2015-219}

\author[a,b]{Pascal du Bosque,}
\emailAdd{dubosque@mpp.mpg.de}
\author[c,d,e]{Falk Hassler,}
\emailAdd{fhassler@unc.edu}
\author[a,b]{and Dieter L\" ust}
\emailAdd{dieter.luest@lmu.de}

\affiliation[a]{Max-Planck-Institut f\"ur Physik\\
F\"ohringer Ring 6, 80805 M\"unchen, Germany}
\affiliation[b]{Arnold-Sommerfeld-Center f\"ur Theoretische Physik\\
Fakult\"at f\"ur Physik, Ludwig-Maximilians-Universit\"at M\"unchen\\
Theresienstra\ss e 37, 80333 M\"unchen, Germany}
\affiliation[c]{University of North Carolina\\Department of Physics and Astronomy\\
Phillips Hall, CB \#3255, 120 E. Cameron Ave., Chapel Hill, NC 27599-3255, USA}
\affiliation[d]{City University of New York\\The Graduate Center\\
365 Fifth Avenue, New York, NY 10016, USA}
\affiliation[e]{Columbia University\\
Department of Physics\\
Pupin Hall, 550 West 120th St., New York, NY 10027, USA}

\abstract{A flux formulation of Double Field Theory on group manifold is derived and applied to study generalized Scherk-Schwarz compactifications, which give rise to a bosonic subsector of half-maximal, electrically gauged supergravities. In contrast to the flux formulation of original DFT, the covariant fluxes split into a fluctuation and a background part. The latter is connected to a $2D$-dimensional, pseudo Riemannian manifold, which is isomorphic to a Lie group embedded into O($D,D$). All fields and parameters of generalized diffeomorphisms are supported on this manifold, whose metric is spanned by the background vielbein $E_A{}^I \in$ GL($2D$). This vielbein takes the role of the twist in conventional generalized Scherk-Schwarz compactifications. By doing so, it solves the long standing problem of constructing an appropriate twist for each solution of the embedding tensor. Using the geometric structure, absent in original DFT, $E_A{}^I$ is identified with the left invariant Maurer-Cartan form on the group manifold, in the same way as it is done in geometric Scherk-Schwarz reductions. We show in detail how the Maurer-Cartan form for semisimple and solvable Lie groups is constructed starting from the Lie algebra. For all compact embeddings in O($3,3$), we calculate $E_A{}^I$.}

\begin{document}
\maketitle

\section{Introduction}
Double Field Theory (DFT) \cite{Siegel:1993th,Hull:2009mi,Hull:2009zb,Hohm:2010jy,Hohm:2010pp, Hohm:2011ex} is an attempt to construct a duality invariant effective string action. One interesting question within this context is, whether this can be done in a background independent way. Another important question is, if DFT always leads back to standard supergravity on geometric spaces, or if it is possible that non-geometric string backgrounds can be consistently included into the DFT framework, after relaxing the strong constraint in one way or the other. Recently a new version, named Double Field Theory on group manifolds or abbreviated by DFT${}_\mathrm{WZW}$, was constructed in \cite{Blumenhagen:2014gva}. This theory was derived using Closed String Field Theory (CSFT) applied to a Wess-Zumino-Witten (WZW) model and the associated Ka\v{c}-Moody current algebras up to cubic order in the fields. Later on, it was reformulated in terms of a generalized metric \cite{Blumenhagen:2015zma} and thereby extrapolated to all orders in the fields. In comparison to original DFT of \cite{Siegel:1993th,Hull:2009mi,Hull:2009zb,Hohm:2010jy,Hohm:2010pp, Hohm:2011ex}, which was derived starting from toroidal backgrounds, it gives rise to additional structures. E.g. it comes with new terms in its action, in its generalized Lie derivative, which mediates the gauge transformations of the theory, and in the strong constraint. Furthermore, the theory possesses a manifest $2D$-diffeomorphism invariance through the consequent use of covariant derivatives. This new symmetry is a consequence of the explicit splitting into background fields and fluctuations emerging in DFT${}_\mathrm{WZW}$. Revoking this splitting by imposing the optional extended strong constraint, the known results of original DFT are reproduced and the $2D$-diffeomorphism invariance is broken \cite{Blumenhagen:2015zma}. While fluctuations in our theory are still governed by a strong constraint, the background only has to fulfill the much weaker Jacobi-Identity. An important result visible in DFT${}_\mathrm{WZW}$ is that this relaxation of the strong constraint for the background is closely related to the closure constraint in the original flux formulation \cite{Geissbuhler:2011mx,Grana:2012rr,Geissbuhler:2013uka,Aldazabal:2013sca} and allows to treat genuinely non-geometric background, which are not T-dual to any geometric configuration.

String theory on a curved background space generally needs the addition of fluxes in order to deal with a conformally invariant string background. In particular for the string propagation on a group manifold, the presence of the $H$-flux is required by conformal invariance. A main objective of this paper is to derive a flux formulation of Double Field Theory on group manifolds and to apply it to study generalized Scherk-Schwarz compactifications of DFT${}_\mathrm{WZW}$. We will see that the flux formulation's action on group manifolds
\begin{equation}\label{eqn:actiondftwzwfluxform}
  S = \int d^{2 D}X \, e^{-2d} \big( S^{\hat A\hat B} \mathcal{F}_{\hat A} \mathcal{F}_{\hat B} + \frac{1}{4} \mathcal{F}_{\hat A\hat C\hat D} \, \mathcal{F}_{\hat B}{}^{\hat C\hat D} \, S^{\hat A\hat B} - \frac{1}{12} \mathcal{F}_{\hat A\hat C\hat E} \, \mathcal{F}_{\hat B\hat D\hat F} \, S^{\hat A\hat B} S^{\hat C\hat D} S^{\hat E\hat F} \big)\,,
\end{equation}
derived in the course of this paper, formally matches the results in original DFT. However, we use a slightly different index convention: Hatted indices $\hat{A}, \ldots, \hat{F}$ are associated to the double Lorentz group. They are converted to O($D,D$) indices by the fluctuation vielbein $\tilde{E}_{\hat{A}}{}^B$, where both flat indices $A, \ldots, F$ and hatted indices $\hat{A}, \ldots, \hat{F}$ run from $1, \ldots, 2D$ (see section~\ref{sec:covfluxesWZW} for details). The covariant fluxes
\begin{equation}
  \mathcal{F}_{\hat A\hat B\hat C} = \tilde F_{\hat A\hat B\hat C} + F_{\hat A\hat B\hat C}
\end{equation}
appearing in \eqref{eqn:actiondftwzwfluxform} are quite different from the original results. They explicitly split into a fluctuation part $\tilde F_{\hat A\hat B\hat C}$ and a background part $F_{\hat A\hat B\hat C}$. While the former is based on an O($D,D$)-valued fluctuation generalized vielbein $\tilde E_{\hat A}{}^B$, which has to fulfill the strong constraint
\begin{equation}
  D_A D^A \cdot = 0\,,
\end{equation}
the latter arises as the structure coefficients of the background group manifold whose tangent space is spanned by the vielbein $E_A{}^I$ $\in$ GL($2D$), where $I,J,K$ also run from $1,\ldots, 2D$:
\begin{align}
  \tilde F_{\hat A\hat B\hat C} = 3 D_{[\hat A} \tilde E_{\hat B}{}^E \tilde E_{\hat C]E} \quad \text{and} \quad
  F_{\hat A\hat B\hat C} &= \tilde E_{\hat A}{}^D \tilde E_{\hat B}{}^E \tilde E_{\hat C}{}^F F_{DEF} \\ \text{with} \quad
  F_{ABC} &= 2 D_{[A} E_{B]}{}^I E_{CJ} \,. \nonumber
\end{align}
Note that we use the flat derivatives
\begin{equation}
  D_A = E_A{}^I \partial_I \quad \text{and} \quad
  D_{\hat A} = E_{\hat A}{}^B D_B\,.
\end{equation}
Moreover, the covariant fluxes transform as scalars under generalized diffeomorphisms and $2D$-diffeomorphisms. Remarkably, the background part is much more flexible than the fluctuation part. It is only restricted by the Jacobi identity
\begin{equation}
  F_{AB}{}^E F_{EC}{}^D + F_{CA}{}^E F_{EB}{}^D + F_{BC}{}^E F_{EA}{}^D = 0
\end{equation}
which is equivalent to the closure constraint \cite{Grana:2012rr,Geissbuhler:2013uka,Aldazabal:2013sca} in the original formulation for constant fluxes. Thus, this splitting allows to treat all possible solutions of the embedding tensor and not only the geometric subset. 

Besides the manifest invariance under generalized and $2D$-diffeomorphisms, the action \eqref{eqn:actiondftwzwfluxform} is also invariant under double Lorentz transformations. Its equations of motion
\begin{equation}
  \mathcal{G} = 0 \quad \text{and} \quad 
  \mathcal{G}^{[\hat A\hat B]} = 0
\end{equation}
have the same form as in the original formulation. The absence of the strong constraint violating term $1/6 F_{ABC} F^{ABC}$ in \eqref{eqn:actiondftwzwfluxform}, proposed by \cite{Geissbuhler:2013uka}, results directly from the CFT origin of the theory. A non-vanishing value of this term would result in a conformal anomaly. Still, this term can be added by hand without spoiling any symmetries in order to reproduce the scalar potential of half-maximal, electrically gauged supergravities.

Especially in order to perform generalized Scherk-Schwarz compactifications \cite{Geissbuhler:2011mx,Hohm:2013nja,Aldazabal:2013sca,Geissbuhler:2013uka}, which recently got a lot of attention in DFT \cite{Berman:2013cli,Dibitetto:2012rk,Berman:2013uda,Blair:2015eba} but also in Exceptional Field Theories (EFTs) \cite{Hohm:2014qga,Baguet:2015sma}, a flux formulation is the preferred starting point. Hence, we directly apply the results obtained in the first part of this paper to discuss these compactifications in our new framework. With an appropriate compactification ansatz, we obtain a bosonic subsector of a half-maximal, electrically gauged supergravity
\begin{align}
  S_\mathrm{eff} = \int d^{D-n}x \sqrt{-g} \, e^{-2\phi} \Big( &R + 4 \partial_\mu \phi \, \partial^\mu \phi - \frac{1}{12} \widehat{G}_{\mu\nu\rho} \widehat{G}^{\mu\nu\rho}  \nonumber \\
  &-\frac{1}{4} \widehat{\mathcal H}_{AB} \widehat{\mathcal F}^{A \mu\nu} \widehat{\mathcal F}^B{}_{\mu\nu} + \frac{1}{8} \widehat{\mathcal D}_\mu \widehat{\mathcal H}_{AB} \widehat{\mathcal D}^\mu \widehat{\mathcal H}^{AB} - V \Big)\label{eqn:lowdimeffactionwzw}
\end{align}
as lower dimensional effective theory. Now, the background vielbein $E_A{}^I$ takes the role of the twist $U_I{}^J$ appearing in generalized Scherk-Schwarz compactifications of original DFT. As it is much less restricted than the twist, e.g. it only has to be a GL($2D$) element instead of being limited to the subgroup O($D,D$), it can be identified with the left-invariant Maurer-Cartan form of the effective theory's gauge group. In original DFT, this possibility is ruled out. Thus, there is no explicit construction of the twists $U_I{}^J$ starting from a solution of the embedding tensor. It has to be `guessed', which of course is an unsatisfactory situation. This problem is solved in DFT${}_\mathrm{WZW}$. Interestingly, these new results are in perfect accordance with standard, geometric Scherk-Schwarz compactifications \cite{Scherk:1978ta,Scherk:1979zr} where the twist is chosen as a Maurer-Cartan form, too.

The paper is organized as follows: In the first part, which is contained in section~\ref{sec:FluxFormWZW}, we successively go through all the steps necessary to rewrite the generalized metric action of DFT${}_\mathrm{WZW}$ in terms of the covariant fluxes presented above. Afterwards we discuss in section~\ref{sec:missingFABCFABCterm} the absence of the strong constraint violating term $1/6 F_{ABC} F^{ABC}$, which was introduced in the original flux formulation to reproduce the scalar potential of half-maximal, electrically gauge supergravities. Furthermore, we prove the double Lorentz invariance of the action in section~\ref{sec:doublelorentz}. Next, the gauge transformations and equations of motions are derived. The second part of the paper in section~\ref{sec:genSSWZW} is dedicated to generalized Scherk-Schwarz compactifications. After a short review of the embedding tensor formalism, especially in $n=3$ internal dimensions, original DFT is discussed. Here, we highlight the problem of constructing the twist, mentioned above. In section~\ref{sec:genSSDFTWZW}, we switch to the new flux formulation of DFT${}_\mathrm{WZW}$. In this framework, the generalized background vielbein $E_A{}^I$ takes the role of the twist and can be chosen as the left invariant Maurer-Cartan form on the group manifold. We present explicitly how to construct it, starting form an arbitrary solution of the embedding tensor, in section~\ref{sec:twistconstr}. Finally, section~\ref{sec:conclusion} concludes the paper. In the appendix, we provide the background generalized vielbeins for all compact embeddings in O($3,3$).

\section{Flux formulation}
\label{sec:FluxFormWZW}
Starting from the generalized metric formulation, which is shortly reviewed in section~\ref{sec:genmetric}, we derive the corresponding flux formulation. To this end, we first identify the covariant fluxes in our framework in section~\ref{sec:covfluxesWZW}. Afterwards, we rewrite the generalized metric action \eqref{eqn:actiongenmetric} in terms of these objects, yielding the desired flux formulation. Moreover, we discuss its symmetries and equations of motion.

\subsection{Review of the generalized metric formulation}\label{sec:genmetric}
In the following, we present a compact review of the DFT${}_\mathrm{WZW}$ generalized metric formulation, derived in \cite{Blumenhagen:2015zma}. It is going to be the starting point for the derivation of the flux formulation in the next sections. The theory is formulated on a $2D$-dimensional space with the coordinates
\begin{equation}
  X^I = \begin{pmatrix} x^i & x^{\bar i} \end{pmatrix}\,.
\end{equation}
Doubled, curved indices are denoted by capital letters beginning from $I$. They run from one to $2D$ and decompose into unbared and bared indices, each of them running from 1 to $D$. Doubled indices are lowered and raised with
\begin{equation}
  \eta_{IJ} = E^A{}_I E^B{}_J \, \eta_{AB} \quad \text{and its inverse} \quad
  \eta^{IJ} = E_A{}^I E_B{}^J \eta^{AB}\,.
\end{equation}
Besides curved indices, also flat indices appear in this context. The latter are represented by letters ranging from $A$ to $H$ and are linked to the former by the generalized background vielbein $E_A{}^I$ and its inverse transpose $E^A{}_I$. In order to explicitly calculate $\eta_{IJ}$, we define its flat version
\begin{equation}
  \eta_{AB} = \begin{pmatrix} \eta_{ab} & 0 \\
    0 & -\eta_{\bar a\bar b}
  \end{pmatrix} \quad \text{and} \quad
  \eta^{AB} = \begin{pmatrix} \eta^{ab} & 0 \\
    0 & -\eta^{\bar a\bar b}
  \end{pmatrix}\,.
\end{equation}
Its constituents $\eta_{ab}$ and $\eta_{\bar a\bar b}$ are both Minkowski metrics with signature $(+,\,-,\,\dots,\,-)$. As opposed to the original DFT framework \cite{Hull:2009mi,Hohm:2010jy,Hohm:2010pp}, the vielbein $E_A{}^I$ is not restricted to be O($D,D$) valued. It is an element of GL($2D$) and generally depends on all coordinates $X^I$. Taking into account the partial derivative
\begin{equation}
  \partial_I = \begin{pmatrix} \partial_i & \partial_{\bar i} \end{pmatrix}
\end{equation}
on the target space, we are able to define the flat derivative
\begin{equation}
  D_A = E_A{}^I \partial_I\,.
\end{equation}
The commutator of two such flat derivatives gives rise to another one, namely
\begin{equation}
  [ D_A, D_B ] = F_{AB}{}^C D_C\,.
\end{equation}
This relation allows to define the structure coefficients
\begin{equation}\label{eqn:FABC&OmegaABC}
  F_{ABC} = 2 \Omega_{[AB]C} \quad \text{with the coefficients of anholonomy} \quad
  \Omega_{ABC} = D_A E_B{}^I E_{C I}\,.
\end{equation}
For DFT${}_\mathrm{WZW}$, they have to be constant and totally antisymmetric, which restricts the doubled background space to group manifolds. In order to write the action and its gauge transformations in a compact form, it is convenient to introduce the covariant derivative
\begin{equation}\label{eqn:flatcovderv}
  \nabla_A V^B = D_A V^B + \frac{1}{3} F^B{}_{AC} V^C\,.
\end{equation}
It possesses the following properties:
\begin{itemize}
  \item Compatibility with the frame
    \begin{equation}
      \nabla_A E_B{}^I = D_A E_B{}^I - \frac{1}{3} F^C{}_{AB} E_C{}^I + E_A{}^J \Gamma^I{}_{JK} E_B{}^K = 0\,,
    \end{equation}
    which allows to calculate the Christoffel symbols
    \begin{equation}
      \Gamma^I{}_{JK} =  \frac{1}{3} F^I{}_{JK} - \Omega_{JK}{}^I\,.
    \end{equation}
  \item Compatibility with the $\eta$ metric
    \begin{equation}
      \nabla_A \, \eta_{BC} = 0\,.
    \end{equation}
  \item Compatibility with integration by parts
    \begin{equation}
      \int d X^{2D} e^{-2 \bar d}\, v (\nabla_A w) = - \int d X^{2D} e^{-2 \bar d}\, (\nabla_A v) w
    \end{equation}
    where $\bar d$ denotes the background generalized dilaton and $v$, $w$ are placeholders for tensorial objects contracting to a scalar. This identity is equivalent to
    \begin{equation}\label{eqn:nablaIe-2bard=0}
      \nabla_I e^{-2 \bar d} = \partial_I e^{-2\bar d} - \Gamma^J{}_{IJ} e^{-2\bar d} = 0
        \quad \text{or} \quad \Omega_{IJ}{}^J = 2 \partial_I \bar d\,,
    \end{equation}
    where we used that $e^{-2 \bar d}$ transforms as a scalar density with weight $+1$.
\end{itemize}
After this prelude, we are able to write down the DFT${}_\mathrm{WZW}$ action
\begin{equation}\label{eqn:actiongenmetric}
  S = \int d^{2n} X e^{-2d} \, \mathcal{R}\,,
\end{equation}
with the generalized curvature scalar
\begin{align}
  \mathcal{R} &= 4 \mathcal{H}^{AB} \nabla_A \nabla_B d - \nabla_A \nabla_B \mathcal{H}^{AB} - 4 \mathcal{H}^{AB} \nabla_A d \, \nabla_B d + 4 \nabla_A d \, \nabla_B \mathcal{H}^{AB} \nonumber \\ &\,+ \frac{1}{8} \mathcal{H}^{CD} \nabla_C \mathcal{H}_{AB} \nabla_D \mathcal{H}^{AB} - \frac{1}{2} \mathcal{H}^{AB} \nabla_B \mathcal{H}^{CD} \nabla_D \mathcal{H}_{AC} + \frac{1}{6} F_{ACD} F_B{}^{CD} \mathcal{H}^{AB}\,,
\label{eqn:gencurvature}
\end{align}
which was derived in \cite{Blumenhagen:2015zma}. Its dynamical fields are the generalized metric, fulfilling
\begin{equation}
  \mathcal{H}^{AC} \mathcal{H}^{BD} \eta_{CD} = \eta_{AB}\,,
\end{equation}
and the generalized dilaton $d$. The action \eqref{eqn:actiongenmetric} is invariant under generalized diffeomorphisms
\begin{align}
  \delta_\xi \mathcal{H}^{AB} = \mathcal{L}_\xi \mathcal{H}^{AB} &= \lambda^C \nabla_C \mathcal{H}^{AB} + (\nabla^A \lambda_C - \nabla_C \lambda^A) \mathcal{H}^{CB} + (\nabla^B \lambda_C - \nabla_C \lambda^B) \mathcal{H}^{AC} \nonumber \\
  \delta_\xi d = \mathcal{L}_\xi d & = \xi^A \nabla_A d - \frac{1}{2} \nabla_A \xi^A\,, \label{eqn:gendiffHAB&tilded}
\end{align}
mediated by the generalized Lie derivative, if the strong constraint
\begin{equation}\label{eqn:SC}
  \nabla_A D^A \cdot = 0
\end{equation}
holds for fluctuations and the background structure coefficients fulfill the Jacobi identity
\begin{equation}\label{eqn:jacobiid}
    F_{AB}{}^E F_{EC}{}^D + F_{CA}{}^E F_{EB}{}^D + F_{BC}{}^E F_{EA}{}^D = 0 \,.
\end{equation}
The placeholder $\cdot$ stands for $\mathcal{H}^{AB}$, $d$, the parameter $\xi^A$ of the gauge transformation $\delta_\xi$ and arbitrary products of them. They have to be treated like scalars in equation~\eqref{eqn:SC}. Thus, the covariant derivative only acts on the index of $D^A$. Imposing both the strong constraint and the Jacobi identity, the commutator of two gauge transformations
\begin{equation}
  [\mathcal{L}_{\xi_1},\, \mathcal{L}_{\xi_2}] = \mathcal{L}_{[\xi_1, \xi_2]_\mathrm{C}}
\end{equation}
gives rise to another gauge transformation. Its resulting parameter is governed by the C-bracket
\begin{equation}\label{eqn:cbracket}
  [\xi_1, \xi_2]_C^A = \xi_1^B \,\nabla_B \xi_2^A - \frac{1}{2} \xi_1^B\, \nabla^A\, \xi_{2\,B} - (1 \leftrightarrow 2)
\end{equation}
and the gauge algebra closes.

Besides generalized diffeomorphisms, the action \eqref{eqn:actiongenmetric} is also manifestly invariant under ordinary $2D$-diffeomorphisms. They are dictated by the Lie derivative $L_\xi$ and the covariant derivative $\nabla_I$ is covariant with respect to them. This additional symmetry is absent in the original generalized metric formulation of DFT. By applying the optional extended strong constraint
\begin{equation}\label{eqn:extendedSC}
  \partial_I b \, \partial^I f = 0\,,
\end{equation}
linking background fields $b$ and fluctuations $f$, and by further restricting the background generalized vielbein to be O($D,D$) valued, one breaks the $2D$-diffeomorphism invariance. In this case, the original formulation emerges as a very special case of DFT${}_\mathrm{WZW}$.

\subsection{Covariant fluxes}\label{sec:covfluxesWZW}
Before writing the DFT${}_\mathrm{WZW}$ action in the flux formulation, we first have to fix its constituents, the covariant fluxes. Therefore, we introduce the composite generalized vielbein
\begin{equation}\label{eqn:fluxepsilon}
  \mathcal{E}_{\hat A}{}^I = \tilde{E}_{\hat A}{}^B {E_B}^I\,,
\end{equation}
which combines the background vielbein $E_A{}^I$ with a new vielbein $\tilde{E}_{\hat A}{}^B$, capturing fluctuations around the background. While the former is not O($D,D$) valued, the latter is and thus fulfills
\begin{equation}
  \eta_{AB} = \tilde{E}^{\hat C}{}_A \, \eta_{\hat C\hat D} \, \tilde{E}^{\hat D}{}_B\,,
\end{equation}
where $\eta_{AB}$ and $\eta_{\hat A\hat B}$ have exactly the same entries. Much more, it allows to express the generalized metric as
\begin{equation}\label{eqn:fluxgenmetric}
  \mathcal{H}_{AB} = \tilde{E}^{\hat C}{}_A \, S_{\hat C\hat D} \, \tilde{E}^{\hat D}{}_B\,.
\end{equation}
It is of great importance to distinguish between the different indices appearing in the different vielbeins. We already encountered the curved indices $I$, $J$, $K$, $\dots$ and their flat counter parts. Now, we also use hatted indices like $\hat A$, $\hat B$, $\hat C$, \ldots\,. As we are going to see shortly, these indices are connected to the doubled Lorentz symmetry, we discuss in section~\ref{sec:doublelorentz}. At the first glance, it seems puzzling to have two different generalized vielbeins, while in the original formulation one is sufficient. The additional structure, introduced by the background generalized vielbein $E_A{}^I$, can be illustrated through the following diagram:
\begin{equation}\label{eqn:groupindices}
  \tikz[baseline]{\matrix (m) [ampersand replacement=\&, matrix of math nodes, column sep=4em] {\text{O}(1,D-1)\times\text{O}(D-1,1) \& \text{O}(D,D) \& \text{GL}(2 D) \\};
    \draw[->] (m-1-3) -- (m-1-2) node[midway, above] {$\eta_{IJ}$} node[midway, below] {$E_B{}^I$};
    \draw[->] (m-1-2) -- (m-1-1) node[midway, above] {$\mathcal{H}_{AB}$} node[midway, below] {${\tilde E}_{\hat A}{}^B$};}\,.
\end{equation}
Starting point is a $2D$-dimensional smooth manifold $M$ equipped with a pseudo Riemannian metric $\eta$, which exhibits a split signature. It reduces the manifold's structure group from GL($2D$) to O($D,D$). The corresponding frame bundle on $M$ is given by the background generalized vielbein $E_A{}^I$. Moreover, there is the generalized metric $\mathcal{H}_{AB}$. It further reduces the structure group to the double Lorentz group O($1,D-1$)$\times$O($D-1,1$) and is represented by the fluctuation frame $\tilde E_{\hat A}{}^B$. In original DFT, the information encoded in $\eta_{IJ}$ is missing.

To get familiar with the new, composite generalized vielbein $\mathcal{E}_{\hat A}{}^I$, we calculate the C-bracket
\begin{align}\label{eqn:fluxcbracketepsilon1}
  \big[ \mathcal{E}{}_{\hat A}, \mathcal{E}{}_{\hat B} \big]_C^J \, \mathcal{E}_{\hat CJ} &=  2 \mathcal{E}_{[\hat A}{}^I \partial_I \mathcal{E}_{\hat B]}{}^J \mathcal{E}_{\hat CJ} - \mathcal{E}_{[\hat A}{}^I \partial^J \mathcal{E}_{\hat B]I} \mathcal{E}_{\hat CJ} + \mathcal{T}^J{}_{IK} \mathcal{E}_{\hat A}{}^I \mathcal{E}_{\hat B}{}^K \mathcal{E}_{\hat CJ} \nonumber\\ 
  &= F_{\hat A\hat B\hat C} +2 D_{[\hat A} \, \tilde{E}_{\hat B]}{}^D \tilde{E}_{\hat CD} - D_{\hat C} \, \tilde{E}_{[\hat B}{}^D \tilde{E}_{\hat A]D}\,.
\end{align}
In the first line, we have applied the generalized torsion \cite{Blumenhagen:2014gva}
\begin{equation}\label{eqn:gentorsion}
  \mathcal{T}^I{}_{JK} = - \Omega^I{}_{[JK]}
\end{equation}
of the covariant derivative $\nabla_I$ to express the C-bracket in terms of partial derivatives instead of covariant derivatives. Similar to the use of $E_A{}^I$ to switch between flat and curved indices, we apply $\tilde{E}_{\hat A}{}^B$ to obtain the structure coefficients
\begin{equation}
  F_{\hat A\hat B\hat C} = \tilde{E}_{\hat A}{}^D \tilde{E}_{\hat B}{}^E \tilde{E}_{\hat C}{}^F F_{DEF}
\end{equation}
in hatted indices. Furthermore, we define the coefficients of anholonomy
\begin{equation}\label{eqn:fluxanholonomy}
  \tilde{\Omega}_{\hat A\hat B\hat C} = \tilde{E}_{\hat A}{}^D D_D \tilde{E}_{\hat B}{}^E \tilde{E}_{\hat CE} = D_{\hat A} \tilde{E}_{\hat B}{}^E \tilde{E}_{\hat CE}
\end{equation}
with
\begin{equation}
  D_{\hat A} = \tilde{E}_{\hat A}{}^B D_B
\end{equation}
for the fluctuations analogous to \eqref{eqn:FABC&OmegaABC}. Due to the fact that the metric $\eta_{AB}$ is constant and therefore can be pulled through flat derivatives, they are antisymmetric in their last two indices:
\begin{equation}
  \tilde{\Omega}_{\hat A\hat B\hat C} = -\tilde{\Omega}_{\hat A\hat C\hat B}\,.
\end{equation}
Finally, we introduce the fluxes
\begin{equation}\label{eqn:fluxfluxes}
  \tilde{F}_{\hat A\hat B\hat C} = 3 \tilde{\Omega}_{[\hat A\hat B\hat C]} =
    \tilde{\Omega}_{\hat A\hat B\hat C} +\tilde{\Omega}_{\hat B\hat C\hat A} + \tilde{\Omega}_{\hat C\hat A\hat B}
\end{equation}
in the same way as they are defined in the flux formulation of original DFT. With these definitions \eqref{eqn:fluxcbracketepsilon1} simplifies to
\begin{equation}
  \label{eqn:fluxcbracketepsilon2}
  \big[ \mathcal{E}{}_{\hat A}, \mathcal{E}{}_{\hat B} \big]_C^M \,\mathcal{E}_{\hat CM} =
    F_{\hat A\hat B\hat C} + 2\tilde{\Omega}_{[\hat A\hat B]\hat C} - \tilde{\Omega}_{\hat C[\hat B\hat A]} = F_{\hat A\hat B\hat C} + \tilde{F}_{\hat A\hat B\hat C} := \mathcal{F}_{\hat A\hat B\hat C}
\end{equation}
and allows us to introduce the covariant fluxes $\mathcal{F}_{\hat A\hat B\hat C}$. They decompose into a background part $F_{\hat A\hat B\hat C}$ and a fluctuation part $\tilde{F}_{\hat A\hat B\hat C}$. An alternative way to construct the covariant fluxes makes use of the generalized Lie derivative
\begin{equation}
  \mathcal{E}_{\hat CM} \, \mathcal{L}_{\mathcal{E}_{\hat A}} \mathcal{E}_{\hat B}{}^{M} =
    \big[ \mathcal{E}_{\hat A}, \mathcal{E}_{\hat B} \big]_C^M \mathcal{E}_{\hat CM} + \frac{1}{2} \nabla^M \big( \mathcal{E}_{\hat AN} \mathcal{E}_{\hat B}{}^N \big) = 
    \big[ \mathcal{E}_{\hat A}, \mathcal{E}_{\hat B} \big]_C^M \mathcal{E}_{\hat CM} =
    \mathcal{F}_{\hat A\hat B\hat C}\,.
\end{equation}
By construction, these fluxes are covariant under generalized diffeomorphisms and $2D$-diffeomor\-phisms. Under both, they transform as scalars.

Besides $\mathcal{F}_{ABC}$, the original flux formulation \cite{Geissbuhler:2011mx,Aldazabal:2011nj,Geissbuhler:2013uka} contains $\mathcal{F}_A$. Its embedding in the DFT${}_\mathrm{WZW}$ framework follows from the definition
\begin{align}
  \mathcal{F}_{\hat A} &= - e^{2d} \mathcal{L}_{\mathcal{E}_{\hat A}} e^{-2d} = 
    - e^{2d} \nabla_B \big( \mathcal{E}_{\hat A}{}^B e^{-2d} \big) =
    \tilde{\Omega}^{\hat B}{}_{\hat B\hat A} + 2 D_{\hat A} \, \tilde{d} - \mathcal{E}_{\hat A}{}^B e^{2 \bar d} \nabla_B e^{-2\bar d} \nonumber\\
  &= 2 D_{\hat A} \, \tilde{d} + \tilde{\Omega}^{\hat B}{}_{\hat B\hat A} = \tilde F_{\hat A}\,.\label{eqn:defFA}
\end{align}
Here, we have applied the decomposition
\begin{equation}
  d = \bar d + \tilde d 
\end{equation}
of the generalized dilaton in a fluctuation and background part $\bar d$ and $\tilde d$. Going from the first to the second line, we make further use of \eqref{eqn:nablaIe-2bard=0}, a direct consequence of the covariant derivative's compatibility with integration by parts. As the covariant fluxes derived in the last paragraph, $\mathcal{F}_{\hat A}$ transforms under generalized and $2D$-diffeomorphisms like a scalar.

\subsection{Action}\label{sec:FluxFormAction}
Now, we are ready to derive the action of the DFT${}_\mathrm{WZW}$ flux formulation. Following \cite{Geissbuhler:2011mx}, we start from the generalized curvature scalar \eqref{eqn:gencurvature} and plug in the generalized metric \eqref{eqn:fluxgenmetric}, expressed in terms of the generalized vielbein $\mathcal{E}_{\hat A}{}^I$.

Let us first calculate the term
\begin{align}\label{eqn:fluxmetrictrafo}
  \nabla_{\hat A} \mathcal{H}^{\hat B\hat C} &= \tilde{E}_{\hat A}{}^A \tilde{E}^{\hat B}{}_B \tilde{E}^{\hat C}{}_C \nabla_{A} \mathcal{H}^{BC} \nonumber \\
  &= \tilde{\Omega}_{\hat A\hat D}{}^{\hat B} S^{\hat D\hat C} + \tilde{\Omega}_{\hat A\hat D}{}^{\hat C} S^{\hat B\hat D} + \frac{1}{3} F^{\hat B}{}_{\hat A\hat D} S^{\hat D\hat C} + \frac{1}{3} F^{\hat C}{}_{\hat A\hat D} S^{\hat B\hat D}
\end{align}
which we are going to need several times in the following calculations. Equipped with this result, we obtain for the first two terms in the second line of \eqref{eqn:gencurvature}
\begin{align}
  \frac{1}{8} \mathcal{H}^{CD} \nabla_C \mathcal{H}_{AB} \nabla_D \mathcal{H}^{AB}
  &= \frac{1}{36} F_{\hat A\hat C\hat D}\, F_{\hat B}{}^{\hat C\hat D} S^{\hat A\hat B} - \frac{1}{36} F_{\hat A\hat C\hat E}\, F_{\hat B\hat D\hat F}\, S^{\hat A\hat B} S^{\hat C\hat D} S^{\hat E\hat F} \nonumber\\
  & \quad + \frac{1}{4} \tilde{\Omega}_{\hat A\hat C\hat D}\, \tilde{\Omega}_{\hat B}{}^{\hat C\hat D} S^{\hat A\hat B} - \frac{1}{4} \tilde{\Omega}_{\hat A\hat C\hat E}\, \tilde{\Omega}_{\hat B\hat D\hat F}\, S^{\hat A\hat B} S^{\hat C\hat D} S^{\hat E\hat F} \nonumber \\
  & \quad + \frac{1}{6} F_{\hat A\hat C\hat D}\, \tilde{\Omega}_{\hat B}{}^{\hat C\hat D} S^{\hat A\hat B} - \frac{1}{6} F_{\hat A\hat C\hat E}  \,\tilde{\Omega}_{\hat B\hat D\hat F}\, S^{\hat A\hat B} S^{\hat C\hat D} S^{\hat E\hat F}
\end{align}
and
\begin{align}
  - \frac{1}{2} \mathcal{H}^{AB} &\nabla_B \mathcal{H}^{CD} \nabla_D \mathcal{H}_{AC} = \frac{1}{18} F_{\hat A\hat C\hat D}\, F_{\hat B}{}^{\hat C\hat D} S^{\hat A\hat B}  - \frac{1}{18} F_{\hat A\hat C\hat E}\, F_{\hat B\hat D\hat F}\, S^{\hat A\hat B} S^{\hat C\hat D} S^{\hat E\hat F} \nonumber \\
  &+ \frac{1}{2} \tilde{\Omega}_{\hat A\hat C\hat E}\, \tilde{\Omega}_{\hat D\hat B\hat F}\, S^{\hat A\hat B} S^{\hat C\hat D} S^{\hat E\hat F} - \frac{1}{2} \tilde{\Omega}_{\hat C\hat A\hat D}\, \tilde{\Omega}_{\hat B}{}^{\hat C\hat D} S^{\hat A\hat B} - \frac{1}{2} \tilde{\Omega}_{\hat A\hat C\hat D}\, \tilde{\Omega}^{\hat C}{}_{\hat B}{}^{\hat D} S^{\hat A\hat B} \nonumber \\ 
  &- \frac{1}{2} \tilde{\Omega}_{\hat C\hat D\hat A}\, \tilde{\Omega}^{\hat D}{}_{\hat B}{}^{\hat C} S^{\hat A\hat B} + \frac{1}{3} F_{\hat A\hat C\hat D}\, \tilde{\Omega}_{\hat B}{}^{\hat C\hat D} S^{\hat A\hat B} - \frac{1}{3} F_{\hat A\hat C\hat E}  \,\tilde{\Omega}_{\hat B\hat D\hat F}\, S^{\hat A\hat B} S^{\hat C\hat D} S^{\hat E\hat F}\,.
\end{align}
The remaining third term in this line yields
\begin{equation}\label{eqn:fluxterm3}
  \frac{1}{6} F_{ACD} F_B{}^{CD} \mathcal{H}^{AB} =
    \frac{1}{6} F_{\hat A\hat C\hat D}\, F_{\hat B}{}^{\hat C\hat D} S^{\hat A\hat B}\,.
\end{equation}
Summing up these three terms and combining appropriate terms into covariant fluxes $\mathcal{F}_{\hat A\hat B\hat C}$, we find
\begin{gather}
  \frac{1}{8} \mathcal{H}^{CD} \nabla_C \mathcal{H}_{AB} \nabla_D \mathcal{H}^{AB} - \frac{1}{2} \mathcal{H}^{AB} \nabla_B \mathcal{H}^{CD} \nabla_D \mathcal{H}_{AC} +\frac{1}{6} F_{ACE} F_{BDF} \mathcal{H}^{AB} \eta^{CD} \eta^{EF} = \nonumber \\ 
  \frac{1}{4} \mathcal{F}_{\hat A\hat C\hat E} \mathcal{F}_{\hat B\hat D\hat F} S^{\hat A\hat B} \eta^{\hat C\hat D} \eta^{\hat E\hat F} - \frac{1}{12} \mathcal{F}_{\hat A\hat C\hat E} \mathcal{F}_{\hat B\hat D\hat F} S^{\hat A\hat B} S^{\hat C\hat D} S^{\hat E\hat F} \nonumber \\ - \frac{1}{2} \tilde{\Omega}_{\hat C\hat D\hat A}\, \tilde{\Omega}^{\hat C\hat D}{}_{\hat B}\, S^{\hat A\hat B} - \tilde{\Omega}_{\hat C\hat D\hat A}\, \tilde{\Omega}^{\hat D}{}_{\hat B}{}^{\hat C} S^{\hat A\hat B} - F_{\hat A\hat C\hat D}\, \tilde{\Omega}^{\hat C\hat D}{}_{\hat B}\, S^{\hat A\hat B}\,. \label{eqn:fluxresults1}
\end{gather}
Except for the last line, this result looks already quite promising. Subsequently, we evaluate the terms in the first line of \eqref{eqn:gencurvature}. They give rise to
\begin{align}\label{eqn:fluxterm4}
  4\mathcal{H}^{AB} \nabla_A \nabla_B d &= 4 S^{\hat A\hat B} D_{\hat A} D_{\hat B} \, \tilde{d} - 4 S^{\hat A\hat B} \, \tilde{\Omega}_{\hat A\hat B}{}^{\hat C} D_{\hat C} \, \tilde{d}\,, \\
  \label{eqn:fluxterm5}
  -4 \mathcal{H}^{AB} \nabla_A d \, \nabla_B d &= -4 S^{\hat A\hat B} D_{\hat A} \tilde{d} \, D_{\hat B} \tilde{d}\,, \\
  \label{eqn:fluxterm6}
  4 \nabla_A d \, \nabla_B \mathcal{H}^{AB} &= - 4 D_{\hat A} \tilde{d} \, \tilde{\Omega}^{\hat C}{}_{\hat C\hat B} \, S^{\hat A\hat B} + 4 S^{\hat A\hat B} \, \tilde{\Omega}_{\hat A\hat B}{}^{\hat C} D_{\hat C} \, \tilde{d} \\
\intertext{and}
  - \nabla_A \nabla_B \mathcal{H}^{AB} &= - S^{\hat A\hat B}\, \tilde{\Omega}^{\hat C}{}_{\hat C\hat A} \, \tilde{\Omega}^{\hat D}{}_{\hat D\hat B} + S^{\hat A\hat B} D_{\hat A} \, \tilde{\Omega}^{\hat C}{}_{\hat C\hat B} \nonumber\\ 
  &\quad + \tilde{\Omega}_{\hat B\hat C}{}^{\hat A} S^{\hat B\hat C} \, \tilde{\Omega}^{\hat D}{}_{\hat D\hat A} - D_{\hat A}\, \tilde{\Omega}_{\hat B\hat C}{}^{\hat A} S^{\hat B\hat C}\,. \label{eqn:fluxterm7}
\end{align}
We rewrite the last two terms of \eqref{eqn:fluxterm7} as
\begin{equation}
  - \tilde{E}_{\hat A}{}^A \tilde{E}_{\hat B}{}^B \big( D_A D_B \tilde{E}_{\hat C}{}^M \big) \tilde{E}^{\hat A}{}_M \, S^{\hat B\hat C} + \tilde{\Omega}_{\hat C\hat D\hat A}\, \tilde{\Omega}^{\hat D}{}_{\hat B}{}^{\hat C} S^{\hat A\hat B}\,,
\end{equation}
while the last term in the first line of this equation yields
\begin{equation}
  - \tilde{E}_{\hat A}{}^A \tilde{E}_{\hat B}{}^B \big( D_A D_B \tilde{E}_{\hat C}{}^M \big) \tilde{E}^{\hat A}{}_M \, S^{\hat B\hat C} - F_{\hat A\hat C\hat D}\, \tilde{\Omega}^{\hat C\hat D}{}_{\hat B}\, S^{\hat A\hat B}\,.
\end{equation}
Combining these two results, we find
\begin{align}
  - \nabla_A \nabla_B \mathcal{H}^{AB} &= - S^{\hat A\hat B}\, \tilde{\Omega}^{\hat C}{}_{\hat C\hat A} \, \tilde{\Omega}^{\hat D}{}_{\hat D\hat B} + 2 S^{\hat A\hat B} D_{\hat A} \, \tilde{\Omega}^{\hat C}{}_{\hat C\hat B} \\
  &+ \tilde{\Omega}_{\hat C\hat D\hat A}\, \tilde{\Omega}^{\hat D}{}_{\hat B}{}^{\hat C} S^{\hat A\hat B} + F_{\hat A\hat C\hat D}\, \tilde{\Omega}^{\hat C\hat D}{}_{\hat B}\, S^{\hat A\hat B}\,. \label{eqn:fluxterm7*}
\end{align}
In total, the terms in the first line of \eqref{eqn:gencurvature} give rise to
\begin{gather}
  4 \mathcal{H}^{AB} \nabla_A \nabla_B d - \nabla_A \nabla_B \mathcal{H}^{AB} - 4 \mathcal{H}^{AB} \nabla_A d\, \nabla_B d + 4 \nabla_A d \,\nabla_B \mathcal{H}^{AB} = \nonumber \\
  \qquad 2 S^{\hat A\hat B} D_{\hat A}\,\mathcal{F}_{\hat B} - S^{\hat A\hat B} \mathcal{F}_{\hat A} \mathcal{F}_{\hat B} + \tilde{\Omega}_{\hat C\hat D\hat A}\, \tilde{\Omega}^{\hat D}{}_{\hat B}{}^{\hat C} S^{\hat A\hat B} + F_{\hat A\hat C\hat D}\, \tilde{\Omega}^{\hat C\hat D}{}_{\hat B}\, S^{\hat A\hat B}\,. \label{eqn:fluxresults2}
\end{gather}
Ultimately, we arrive at
\begin{align}\label{eqn:fluxresults3}
  \mathcal{R} =& \frac{1}{4} \mathcal{F}_{\hat A\hat C\hat D} \mathcal{F}_{\hat B}{}^{\hat C\hat D} S^{\hat A\hat B} - \frac{1}{12} \mathcal{F}_{\hat A\hat C\hat E} \mathcal{F}_{\hat B\hat D\hat F}  S^{\hat A\hat B} S^{\hat C\hat D} S^{\hat E\hat F} \nonumber \\ 
  & - \frac{1}{2} \tilde{\Omega}_{\hat C\hat D\hat A}\, \tilde{\Omega}^{\hat C\hat D}{}_{\hat B}\, S^{\hat A\hat B} + 2 S^{\hat A\hat B} D_{\hat A}\,\mathcal{F}_{\hat B} - S^{\hat A\hat B} \mathcal{F}_{\hat A} \mathcal{F}_{\hat B}
\end{align}
by taking \eqref{eqn:fluxresults1} and \eqref{eqn:fluxresults2} into account. Moreover, applying the strong constraint
\begin{equation}
  D_{\hat C} \tilde{E}_{\hat D}{}^A  D^{\hat C} \tilde{E}^{\hat D}{}_B = 0
\end{equation}
for fluctuations, the first term in the second line vanishes. Analogous to the generalized metric formulation of DFT${}_\mathrm{WZW}$ discussed in section~\ref{sec:genmetric}, the strong constraint only is required for fluctuations. For the background, captured by $F_{\hat A\hat B\hat C}$, only the Jacobi identity \eqref{eqn:jacobiid} has to hold. Performing integration by parts
\begin{equation}\label{eqn:ibpDhatA}
  \int d^{2 D}X \, e^{-2d} D_{\hat A} v \, w = \int d^{2 D}X\, ( \mathcal{F}_{\hat A} v \, w - v\, D_{\hat A} w )\,,
\end{equation}
we obtain the action
\begin{align}\label{eqn:actionfluxform}
  S = \int d^{2 D}X \, e^{-2d} \big( S^{\hat A\hat B} \mathcal{F}_{\hat A} \mathcal{F}_{\hat B} + \frac{1}{4} \mathcal{F}_{\hat A\hat C\hat D} \, \mathcal{F}_{\hat B}{}^{\hat C\hat D} \, S^{\hat A\hat B} - \frac{1}{12} \mathcal{F}_{\hat A\hat C\hat E} \, \mathcal{F}_{\hat B\hat D\hat F} \, S^{\hat A\hat B} S^{\hat C\hat D} S^{\hat E\hat F} \big)\,. 
\end{align}
It is manifestly invariant under generalized diffeomorphisms and $2D$-diffeomorphisms, because it only contains covariant fluxes and no additional flat derivatives. Its form is equivalent to the original flux formulation of \cite{Geissbuhler:2013uka} without strong constraint violation terms. We explain in the next section why these terms are absent here. However, the covariant fluxes $\mathcal{F}_{\hat A\hat B\hat C}$ differ significantly from the previous results. They now exhibit an explicit splitting into a fluctuation and a background part. 

In order to demonstrate the transition to the original formulation after imposing the extended strong constraint \eqref{eqn:extendedSC} and restricting the background generalized vielbein to O($D,D$), this splitting has to vanish. Hence, if we remember that imposing these two optional constraints allows us to replace \cite{Blumenhagen:2015zma}
\begin{equation}
  F_{ABC} = 2\Omega_{[AB]C} \quad \text{with} \quad F_{ABC} = 3\Omega_{[ABC]}\,,
\end{equation}
which yields
\begin{equation}
  \mathcal{F}_{\hat A\hat B\hat C} = 3 ( \tilde \Omega_{[\hat A\hat B\hat C]} + \Omega_{[\hat A\hat B\hat C]} ) =
    3 D_{[\hat A} \mathcal{E}_{\hat B}{}^I \mathcal{E}_{\hat C] I}\,.
\end{equation}
This breaks the strict distinction between background and fluctuations. Only the O($D,D$) valued composite vielbein remains. Of course, its dynamics are still governed by the action~\eqref{eqn:actionfluxform}.

\subsubsection{Strong constraint violating terms}\label{sec:missingFABCFABCterm}
The action \eqref{eqn:actionfluxform} reproduces all terms of the original flux formulation \cite{Geissbuhler:2013uka}
\begin{align}
  S_\mathrm{DFT} = \int d^{2D} X\, e^{-2d} &\big( \mathcal{F}_A \mathcal{F}_B S^{AB} + \frac{1}{4} \mathcal{F}_{ACD} \mathcal{F}_B{}^{CD} S^{AB} - \frac{1}{12} \mathcal{F}_{ABC} \mathcal{F}_{DEF} S^{AD} S^{BE} S^{CF} \nonumber \\
    & \quad - \frac{1}{6} \mathcal{F}_{ABC} \mathcal{F}^{ABC} - \mathcal{F}_A \mathcal{F}^A \big)\,,
\end{align}
except for the strong constraint violating ones in the second line. All fluctuations are required to fulfill the strong constraint. Thus, they do not contribute to these missing terms. Nonetheless, one would expect to find at least background contributions of the form
\begin{equation}\label{eqn:candiateterms}
  F_A F^A \quad \text{or} \quad \frac{1}{6} F_{ABC} F^{ABC}\,.
\end{equation}
In order to see why these terms are not appearing either, we go back to the CSFT origins of DFT${}_\mathrm{WZW}$. We only considered CFTs with a constant dilaton. Thus, $F_A = 0$ has to hold and the first term in \eqref{eqn:candiateterms} drops out. Further, remember the expression for the central charge \cite{Blumenhagen:2014gva}
\begin{equation}\label{eqn:centralcharge}
  c  = \frac{k D}{k + h^\vee}
\end{equation}
of the closed strings left moving part, with the level $k$ and the dual Coxeter number $h^\vee$. It gives rise to the total central charge
\begin{equation}\label{eqn:ctotleft}
  c_\mathrm{tot} = c + c_\mathrm{gh} = D - \frac{D h^\vee}{k} + c_\mathrm{gh} + \mathcal{O}(k^{-1})\,,
\end{equation}
after adding the ghost contribution $c_\mathrm{gh}$. Terms of order $k^{-2}$ and higher were excluded during the derivation of DFT${}_\mathrm{WZW}$. Therefore, we also neglect them when computing the central charge. Using 
\begin{equation}
  \eta_{ab} =  -\frac{\alpha' k}{4 h^\vee} F_{ad}{}^c F_{bc}{}^d \,,
\end{equation}
as it was defined in \cite{Blumenhagen:2014gva}, we express the second term in \eqref{eqn:ctotleft},
\begin{equation}
  -\frac{D h^\vee}{k} = \frac{\alpha'}{4} F_{ad}{}^c F_{bc}{}^d \eta^{ab}\,,
\end{equation}
through the unbared structure coefficients\footnote{Note that this identification only works for semisimple Lie algebras whose Killing form is non-degenerate. But this was also exactly the assumption while deriving DFT${}_\mathrm{WZW}$ via CSFT.}. Keeping in mind that the same relations hold for the central charge of the anti-chiral, right moving part, we obtain
\begin{equation}
  c_\mathrm{tot} - \bar c_\mathrm{tot} = \frac{\alpha'}{4} \big( F_{ad}{}^c F_{bc}{}^d \eta^{ab} - F_{\bar{a}\bar{d}}{}^{\bar{c}} F_{\bar{b}\bar{c}}{}^{\bar{d}} \eta^{\bar{a}\bar{b}} \big) = - \frac{\alpha'}{2} F_{ABC} F^{ABC}
\end{equation}
when remembering the decompositions
\begin{equation}
  \eta^{AB} = \frac{1}{2} \begin{pmatrix}
    \eta^{ab} & 0 \\
    0 & -\eta^{\bar a\bar b}
  \end{pmatrix}
  \quad \text{and} \quad
  F_{AB}{}^C = \begin{cases}
    F_{ab}{}^c & \\
    F_{\bar a\bar b}{}^{\bar c} & \\
    0 & \text{otherwise\,.}
  \end{cases}
\end{equation}
This result is proportional to the second term in \eqref{eqn:candiateterms}. As CSFT derivations require that both total central charges $c_\mathrm{tot}$ and $\bar c_\mathrm{tot}$ vanish independently, it has to vanish, too. Another interesting effect of this observation is that the scalar curvature
\begin{equation}
  R = \frac{2}{9} F_{ABC} F^{ABC} = R_{ABC}{}^B \eta^{AC} = 0\,,
\end{equation}
which arises from the Riemann curvature tensor
\begin{equation}
  R_{ABC}{}^D = \frac{2}{9} F_{AB}{}^E F_{EC}{}^D\,,
\end{equation}
induced by the covariant derivative $\nabla_A$, has to vanish.

\subsubsection{Double Lorentz symmetry}\label{sec:doublelorentz}
Besides generalized and 2D-diffeomorphisms invariance, there is local double Lorentz symmetry. It acts on hatted indices, as the one of the fluctuation generalized vielbein, by
\begin{equation}\label{eqn:localodxodsym}
  \tilde E_{\hat A}{}^B \rightarrow T_{\hat A}{}^{\hat C} \tilde E_{\hat C}{}^B
\end{equation}
where the tensor $T_{\hat A}{}^{\hat B}$ has to fulfill the properties
\begin{equation}
  T_{\hat A}{}^{\hat C} \eta_{\hat C\hat D} T_{\hat B}{}^{\hat D} = \eta_{\hat A\hat B}
    \quad \text{and} \quad
  T_{\hat A}{}^{\hat C} S_{\hat C\hat D} T_{\hat B}{}^{\hat D} = S_{\hat A\hat B}\,.
\end{equation}
Whereas in the generalized metric formulation local double Lorentz symmetry is manifest, because there are no hatted indices, in the flux formulation it is not and we have to check it explicitly. To this end, we consider the infinitesimal version of \eqref{eqn:localodxodsym}. We denote such transformations by
\begin{equation}
  \delta_\Lambda \mathcal{E}_{\hat A}{}^I = \Lambda_{\hat A}{}^{\hat B} \mathcal{E}_{\hat B}{}^I\,.
\end{equation}
Furthermore, as a generator of a doubled Lorentz transformations, $\Lambda_{\hat A\hat B}$ fulfills the identities
\begin{equation}\label{eqn:odxodlambda}
  \Lambda_{\hat A\hat B} = - \Lambda_{\hat B\hat A} \quad \text{and} \quad
  \Lambda_{\hat A\hat B} = S_{\hat A\hat C} \Lambda^{\hat C\hat D} S_{\hat D\hat B}\,.
\end{equation}
A short calculation gives rise to the transformation behavior
\begin{align}
  \delta_\Lambda \mathcal{F}_{\hat A\hat B\hat C} &= 3 \big( D_{\hat{[A}} \Lambda_{\hat B\hat C]} + \Lambda_{[\hat A}{}^{\hat D} \mathcal{F}_{\hat B\hat C]\hat D} \big) \\
  \delta_\Lambda \mathcal{F}_{\hat A} &= D^{\hat B} \Lambda_{\hat B\hat A} + \Lambda_{\hat A}{}^{\hat B} \mathcal{F}_{\hat B}
\end{align}
of the covariant fluxes. Note that the last terms in both equations spoil covariance under double Lorentz transformations. Using these results, it is straightforward to calculate
\begin{equation}\label{eqn:odxodaction2}
  \delta_\Lambda S = - \int d^{2n} X \, e^{-2d} \Lambda_{\hat A}{}^{\hat C} \delta^{\hat A\hat B} \mathcal{Z}_{\hat B\hat C}
\end{equation}
with
\begin{equation}\label{eqn:odxodz1}
  \mathcal{Z}_{\hat A\hat B} = D^{\hat C} \mathcal{F}_{\hat C\hat A\hat B} + 2 D_{[\hat A} \mathcal{F}_{\hat B]} - \mathcal{F}^{\hat C} \mathcal{F}_{\hat C\hat A\hat B}\,.
\end{equation}
We do not present the intermediate steps of this calculation, since they are analogous to the derivation for the flux formulation of original DFT \cite{Geissbuhler:2013uka}. For evaluation of $\mathcal{Z}_{\hat A\hat B}$, we split the covariant fluxes $\mathcal{F}_{\hat A\hat B\hat C}$ into their fluctuation and background parts according to \eqref{eqn:fluxcbracketepsilon2}. Consequently, we have to calculate the terms
\begin{align}
  D^{\hat C} \tilde{F}_{\hat C\hat A\hat B} &= D^C \big( D_C \tilde{E}_{[\hat A}{}^D \tilde{E}_{\hat B]D} \big) + \tilde{\Omega}^{\hat C}{}_{\hat C\hat D} \, \tilde{\Omega}^{\hat D}{}_{\hat A\hat B} + \underline{2 D^{\hat C} \tilde{\Omega}_{[\hat A\hat B]\hat C}} \nonumber \\
  D^{\hat C} F_{\hat C\hat A\hat B} &= \tilde{E}_{\hat A}{}^A \tilde{E}_{\hat B}{}^B D^C F_{CAB} + \tilde{\Omega}^{\hat D}{}_{\hat D}{}^{\hat C} F_{\hat C\hat A\hat B} + \underline{2 F_{[\hat A\hat C\hat D} \tilde{\Omega}^{\hat C\hat D}{}_{\hat B]}} \nonumber \\
  2 D_{[\hat A} \tilde F_{\hat B]} &= 2 F_{\hat A\hat B}{}^{\hat C} D_{\hat C} \, \tilde{d} + 4 \tilde{\Omega}_{[\hat A\hat B]}{}^{\hat C} D_{\hat C} \, \tilde{d} + \underline{2 D_{[\hat A} \tilde{\Omega}^{\hat C}{}_{\hat C\hat B]}} \nonumber \\
  - \tilde F^{\hat C} F_{\hat C\hat A\hat B} &= - 2 F_{\hat A\hat B}{}^{\hat C} D_{\hat C} \, \tilde{d} - \tilde{\Omega}^{\hat D}{}_{\hat D}{}^{\hat C} F_{\hat C\hat A\hat B} \nonumber \\
  - \tilde F^{\hat C} \tilde{F}_{\hat C\hat A\hat B} &= - 2 \tilde{\Omega}^{\hat C}{}_{\hat A\hat B} D_{\hat C} \, \tilde{d} - 4 \tilde{\Omega}_{[\hat A\hat B]}{}^{\hat C} D_{\hat C} \, \tilde{d} - \tilde{\Omega}^{\hat D}{}_{\hat D}{}^{\hat C} \tilde{\Omega}_{\hat C\hat A\hat B} - \underline{2 \tilde{\Omega}^{\hat D}{}_{\hat D}{}^{\hat C} \tilde{\Omega}_{[\hat A\hat B]\hat C}} \nonumber\,.
\end{align}
The underlined terms cancel due to the identity
\begin{equation}
  2 D^{\hat C} \tilde{\Omega}_{[\hat A\hat B]\hat C} - 2 \tilde{\Omega}^{\hat D}{}_{\hat D}{}^{\hat C} \, \tilde{\Omega}_{[\hat A\hat B]\hat C} =
    - 2 F_{[\hat A\hat C\hat D} \tilde{\Omega}^{\hat C\hat D}{}_{\hat B]} - 2 D_{[\hat A} \tilde{\Omega}^{\hat C}{}_{\hat C\hat B]}
\end{equation}
which arises after swapping two flat derivatives. Thus, equation \eqref{eqn:odxodz1} yields
\begin{equation}\label{eqn:odxodz2}
  \mathcal{Z}_{\hat A\hat B} = D^C \big( D_C \tilde{E}_{[\hat A}{}^D \tilde{E}_{\hat B]D} \big) -2 \tilde{\Omega}^{\hat C}{}_{\hat A\hat B} D_{\hat C} \, \tilde{d} + \tilde{E}_{\hat A}{}^A \tilde{E}_{\hat B}{}^B D^C F_{CAB}\,,
\end{equation}
where the first two terms vanish under the strong constraint. The remaining term gives rise to
\begin{equation}\label{eqn:odxodxz3}
  \mathcal{Z}_{\hat A\hat B} = \tilde{E}_{\hat A}{}^A \tilde{E}_{\hat B}{}^B D^C F_{CAB}\,.
\end{equation}
The structure coefficients $F_{ABC}$ are constant, as we are on a group manifold, and we finally find $\mathcal{Z}_{\hat A\hat B}=0$. Hence, we have proved the invariance of the action \eqref{eqn:actionfluxform} under double Lorentz transformations.

\subsection{Gauge transformations}\label{sec:gaugetrafo}
In the flux formulation it is convenient to write all quantities in hatted indices. Thus, we now check how the gauge transformations \eqref{eqn:gendiffHAB&tilded} of DFT${}_\mathrm{WZW}$ act on these indices. Therefore, we introduce an arbitrary vector in the canonical way
\begin{equation}\label{eqn:hatsplitting}
  V^{\hat A} = \tilde{E}^{\hat A}{}_B V^B\,.
\end{equation}
Parameters of a gauge transformation are vectors, too. Hence, they are given in the same fashion. This splitting allows to evaluate the generalized Lie derivatives as
\begin{align}\label{eqn:hatliederivative}
  \mathcal{L}_{\xi} V^{\hat A} &= \xi^{\hat B} D_{\hat B} V^{\hat A} +  \big( D^{\hat A} \xi_{\hat B} - D_{\hat B} \xi^{\hat A} \big) V^{\hat B} + \mathcal{F}^{\hat A}{}_{\hat B\hat C} \, \xi^{\hat B} V^{\hat C} \quad \text{and} \\
  \mathcal{L}_{\xi} d &= \frac{1}{2} \xi^{\hat A} \mathcal{F}_{\hat A} - \frac{1}{2} D_{\hat A} \xi^{\hat A}\,,
\end{align}
where $\mathcal{F}_{\hat A\hat B\hat C}$ and $\mathcal{F}_{\hat A}$ denote the covariant fluxes defined in \eqref{eqn:fluxcbracketepsilon2} and \eqref{eqn:defFA}. Note that this result formally matches with the original flux formulation. But, as for the action, the covariant fluxes are defined differently and split into a background and a fluctuation part.

Furthermore, equipped with \eqref{eqn:hatsplitting}, we are also able to compute the C-bracket \eqref{eqn:cbracket} in hatted indices. Doing so, we obtain
\begin{equation}\label{eqn:hatcbracket}
  \big[ \xi_1,\xi_2 \big]_C^{\hat A} = \xi_1^{\hat B} D_{\hat B} \xi_2^{\hat A} - \frac{1}{2} \xi_1^{\hat B} D^{\hat A} \xi_{2\,\hat B} + \frac{1}{2}\mathcal{F}^{\hat A}{}_{\hat B\hat C} \xi_1^{\hat B} \xi_2^{\hat C} - (1 \leftrightarrow 2)\,.
\end{equation}
Again, the same comments as for the action and the generalized Lie derivative hold.

\subsection{Equations of motion}\label{sec:eom}
Now, we derive the equations of motion, following \cite{Aldazabal:2013sca, Geissbuhler:2013uka}. The variations of the action \eqref{eqn:actionfluxform} with respect to the dilaton fluctuations $\tilde d$ and the fluctuation vielbein $\tilde{E}_{\hat A}{}^B$ can be formally written as
\begin{equation}\label{eqn:vardtilde}
  \delta_{\tilde d} \, S = \int d^{2n} X  e^{-2d} \, \mathcal{G} \, \delta\tilde{d}
\end{equation}
and
\begin{equation}\label{eqn:varEtilde}
  \delta_{E} \, S = \int d^{2n} X  e^{-2d} \, \mathcal{G}^{\hat A\hat B} \, \delta\tilde{E}_{\hat A\hat B} 
    \quad \text{with} \quad
  \delta\tilde{E}_{\hat A\hat B} = \delta\tilde{E}_{\hat A}{}^C \tilde{E}_{\hat BC}\,.
\end{equation}
Because $\delta\tilde{E}_{\hat A\hat B}$ is antisymmetric, which immediately follows from
\begin{equation}
  \delta ( \tilde E_{\hat A}{}^C \tilde E_{\hat B C} ) = \delta \eta_{\hat A\hat B} = 0\,,
\end{equation}
only the antisymmetric part of $\mathcal{G}^{\hat A\hat B}$ contributes. Evaluating the variations \eqref{eqn:vardtilde} and \eqref{eqn:varEtilde} explicitly, we find
\begin{equation}
  \mathcal{G} = - 2 \mathcal{R} 
    \quad \text{and} \quad
  \mathcal{G}^{[\hat A\hat B]} = 2 S^{\hat D[\hat A} D^{\hat B]} \mathcal{F}_{\hat D} + \big( \mathcal{F}_{\hat D} - D_{\hat D} \big) \check{\mathcal{F}} ^{\hat D[\hat A\hat B]} + \check{\mathcal{F}} ^{\hat C\hat D[\hat A} \mathcal{F}_{\hat C\hat D}{}^{\hat B]}
\end{equation}
with
\begin{equation}
  \check{\mathcal{F}}^{\hat A\hat C\hat E} = \Big( - \frac{1}{2} S^{\hat A\hat B}S^{\hat C\hat D}S^{\hat E\hat F} + \frac{1}{2} S^{\hat A\hat B}\eta^{\hat C\hat D}\eta^{\hat E\hat F} + \frac{1}{2} \eta^{\hat A\hat B}S^{\hat C\hat D}\eta^{\hat E\hat F} + \frac{1}{2} \eta^{\hat A\hat B}\eta^{\hat C\hat D}S^{\hat E\hat F} \Big) \mathcal{F}_{\hat B\hat D\hat F}\,.
\end{equation}
Thus, the equation of motion read
\begin{equation}
  \mathcal{G} = 0 \quad \text{and} \quad 
  \mathcal{G}^{[\hat A\hat B]} = 0\,.
\end{equation}
Again, this result matches the one for the original flux formulation, which was derived in \cite{Geissbuhler:2013uka}. However, keep in mind that the covariant fluxes used here differ significantly from the original ones. The object $\check{\mathcal{F}}^{\hat A\hat B\hat C}$ seems at first glance quite artificial. Its role becomes more obvious, if we rewrite it through
\begin{equation}
  \check{\mathcal{F}}_{\hat A\hat C\hat E} = \mathcal{P}_{\hat A\hat C\hat E}{}^{\hat B\hat D\hat F} \mathcal{F}_{\hat B\hat D\hat F}\,,
\end{equation}
where $\mathcal{P}_{\hat A\hat C\hat E}{}^{\hat B\hat D\hat F}$ incorporates eight different projections
\begin{align}
\mathcal{P}_{\hat A\hat C\hat E}{}^{\hat B\hat D\hat F} =&\;\;\;\:\, P_{\hat A}{}^{\hat B} P_{\hat C}{}^{\hat D} P_{\hat E}{}^{\hat F} - \bar{P}_{\hat A}{}^{\hat B}\bar{P}_{\hat C}{}^{\hat D}\bar{P}_{\hat E}{}^{\hat F}
+ \bar{P}_{\hat A}{}^{\hat B} P_{\hat C}{}^{\hat D} P_{\hat E}{}^{\hat F} + P_{\hat A}{}^{\hat B} \bar{P}_{\hat C}{}^{\hat D} P_{\hat E}{}^{\hat F} \\ &+ P_{\hat A}{}^{\hat B} P_{\hat C}{}^{\hat D} \bar{P}_{\hat E}{}^{\hat F} - \bar{P}_{\hat A}{}^{\hat B} \bar{P}_{\hat C}{}^{\hat D} P_{\hat E}{}^{\hat F} - \bar{P}_{\hat A}{}^{\hat B} P_{\hat C}{}^{\hat D} \bar{P}_{\hat E}{}^{\hat F} - P_{\hat A}{}^{\hat B} \bar{P}_{\hat C}{}^{\hat D} \bar{P}_{\hat E}{}^{\hat F} \nonumber
\end{align}
with the projectors
\begin{equation}
  P_{\hat A}{}^{\hat B} = \frac{1}{2} \big( S_{\hat A}{}^{\hat B} + \delta_{\hat A}{}^{\hat B} \big) \quad\text{and}\quad \bar{P}_{\hat A}{}^{\hat B} = -\frac{1}{2} \big( S_{\hat A}{}^{\hat B} - \delta_{\hat A}{}^{\hat B} \big)\,.
\end{equation}
These two projectors are well known from the equations of motion in the generalized metric formulation \cite{Hohm:2010pp,Blumenhagen:2015zma}.

\section{Generalized Scherk-Schwarz compactification}\label{sec:genSSWZW}
The flux formulation derived in section~\ref{sec:FluxFormWZW} allows us to connect DFT${}_\mathrm{WZW}$ with generalized Scherk-Schwarz compactifications. Evidences for this link were already mentioned in \cite{Blumenhagen:2014gva,Blumenhagen:2015zma}. Here, we make it manifest by applying a slightly adapted generalized Scherk-Schwarz ansatz and derive the low-energy, effective theory in section~\ref{sec:genSSWZW}. As expected, this theory describes a bosonic subsector of a half-maximal, electrically gauged supergravity. All emerging gauged supergravities can be classified in terms of the embedding tensor which is reviewed in section~\ref{sec:embeddingtensor}. Following \cite{Dibitetto:2012rk,Dibitetto:2015bia}, explicit solutions are discussed for compactifications with $n=3$ internal dimensions. Before we present our new results, we shortly review generalized Scherk-Schwarz compactifications in original DFT, where the construction of the twist, which captures all the properties of the compactification, is problematic. In general, the original DFT description is lacking an explicit algorithm to obtain the twist from a solution of the embedding tensor and so one has to start guessing. Thus, it is not clear whether there exist twists for all solutions of the embedding tensor at all. With the results presented in this section, we are now able to evade these problems completely. Hence, we give a detailed prescription to derive the background generalized vielbeins, which take the role of the twist, for arbitrary solutions of the embedding tensor in section~\ref{sec:twistconstr}.

\subsection{Embedding tensor}\label{sec:embeddingtensor}
Before starting with the actual generalized Scherk-Schwarz compactifications, we discuss an essential tool to classify maximal/half-maximal gauged supergravities which arise from these compactifications. This tool is called the embedding tensor $\Theta_I{}^{\alpha}$. For a comprehensive review see e.g.  \cite{Samtleben:2008pe}. It describes the embedding of the supergravity's gauge group into the global symmetry group of the ungauged theory. For DFT, we are interested in embeddings in O($D,D$), the T-duality group of a $D$-dimensional torus. There is a direct relation between the embedding tensor and the structure coefficients
\begin{equation}
\label{eqn:embeddingtensor}
  F_{AB}{}^C = \Theta_A{}^{\alpha} \big( t_{\alpha} \big)_B{}^C = \big(X_A\big)_B{}^C\,,
\end{equation}
of the Lie algebra, related to the gauge group. Here, $t_{\alpha}$ labels the different O($D,D$) generators. Their vector representation, acting on arbitrary doubled vectors $V^A$ as
\begin{equation}
  t_\alpha V^A = V^B (t_\alpha)_B{}^A\,,
\end{equation}
is denoted by $\big(t_{\alpha}\big)_B{}^C$. In general, the embedding tensor must fulfill two conditions: a linear and a quadratic constraint. Each solution to both of them specifies a consistent gauged supergravity.

For higher dimensions, solving these constraints is very challenging. Thus, we here restrict the discussion to $n=3$ internal dimensions. Following \cite{Dibitetto:2012rk,Dibitetto:2015bia}, we are going to find in total twelve different solutions, each of them possessing a continuous parameter $\alpha$. To be more specific, $t_\alpha$ in \eqref{eqn:embeddingtensor} is assumed to describe the six different $\mathfrak{o}(3,3)$ generators. Their vector representation carries indices $A, B, \cdots$ running from $1, \dots, 6$. Group-theoretically, the embedding tensor product lives in the tensor product
\begin{equation}\label{eqn:embtensordecomp}
  \mathbf{6} \otimes \mathbf{15} = \mathbf{6} \oplus \mathbf{\overline{10}} \oplus \mathbf{10} \oplus \mathbf{64}\,,
\end{equation}
where the first factor represents the vector representation, while the second one stands for the adjoint representation labeled by the subscript $\alpha$ in $t_\alpha$. The linear constraint projects out certain irreps. In our case, we only keep the irreps $\mathbf{\overline{10}} \oplus \mathbf{10}$ of the decomposition \eqref{eqn:embtensordecomp}. All other components of the embedding tensor are set to zero. Now, $F_{ABC}$ is in one to one correspondence with the vacuum expectation value (or background part) of the covariant fluxes $\mathcal{F}_{ABC}$, which have exactly the right number ($6\cdot 5 \cdot 4 / 3! = 20$) of independent components.

Following \cite{Dibitetto:2012rk}, we can express $\big(X_A\big)_B{}^C$ through irreps of $\mathfrak{sl}(4)$ instead of using $\mathfrak{so}(3,3)$. Both algebras are isomorphic and the decomposition \eqref{eqn:embtensordecomp} does not change. In order to distinguish between the two different algebras we introduce the fundamental $\mathfrak{sl}(4)$ indices $p,q,r =1,\dots,4$. The relevant  $\mathbf{\overline{10}} \oplus \mathbf{10}$ part of the embedding tensor then reads \cite{Dibitetto:2012rk}
\begin{equation}\label{eqn:embsl41}
  \big(X_{mn}\big)_p{}^q = \frac{1}{2} \delta^q{}_{[m} M_{n]p} - \frac{1}{4}  \varepsilon_{mnpr} \tilde{M}^{rq}
\end{equation}
where $M_{np}$ and $\tilde{M}^{rq}$ are symmetric matrices and $\varepsilon$ labels the Levi-Civita symbol in $4$-dimensions. These symmetric matrices have $4 \cdot 5 / 2 = 10$ independent components each. Thus, we identify $M_{pq}$ with the irrep $\mathbf{10}$, while $\tilde M^{rp}$ lives in the dual irrep $\mathbf{\overline{10}}$. Furthermore, the indices $m$ and $n$ in $\big(X_{mn}\big)_p{}^q$ are antisymmetric and label the $4 \cdot 3 / 2 = 6$ independent components of the $\mathfrak{sl}(4)$ irrep $\mathbf{6}$. The dual representation with two upper antisymmetric indices is given by
\begin{equation}
  X_{mn} = \frac{1}{2} \varepsilon_{mnpq} X^{pq}\,.
\end{equation}

The irreps $\mathbf{\overline{10}} \oplus \mathbf{10}$ are embedded into the product $\mathbf{6} \otimes \mathbf{15}$ by equation \eqref{eqn:embsl41}. However, the structure coefficients live as rank 3 tensor in $\mathbf{6} \otimes \mathbf{6} \otimes \mathbf{6}$. Therefore, $\big(X_{mn}\big)_p{}^q$ needs to be embedded into this product through the relation
\begin{equation}
  \label{eqn:embsl42}
  \big(X_{mn}\big)_{pq}{}^{rs} = 2 \big( X_{mn} \big)_{[p}{}^{[r} \delta_{q]}{}^{s]}\,.
\end{equation}
Finally, we have to go back from $\mathfrak{sl}(4)$ to $\mathfrak{so}(3,3)$. To this end, the irrep $\mathbf{6}$ of the former is related to the latter one by the 't Hooft symbols $\big( G_A \big)^{mn}$. For $n=3$, they read
\begin{align}
  \big(G_1\big)^{mn} &= \frac{1}{\sqrt{2}} \begin{pmatrix}
    0 && -1 && 0 && \phantom{-}0 \\
    1 && \phantom{-}0 && 0 && \phantom{-}0 \\
    0 && \phantom{-}0 && 0 && -1 \\
    0 && \phantom{-}0 && 1 && \phantom{-}0
  \end{pmatrix}\,, &
  \big(G_2\big)^{mn} &= \frac{1}{\sqrt{2}} \begin{pmatrix}
    0 && \phantom{-}0 && -1 && 0 \\
    0 && \phantom{-}0 && \phantom{-}0 && 1 \\
    1 && \phantom{-}0 && \phantom{-}0 && 0 \\
    0 && -1 && \phantom{-}0 && 0
  \end{pmatrix}\,, \nonumber \\
  \big(G_3\big)^{mn} &= \frac{1}{\sqrt{2}} \begin{pmatrix}
    0 && 0 && \phantom{-}0 && -1 \\
    0 && 0 && -1 && \phantom{-}0 \\
    0 && 1 && \phantom{-}0 && \phantom{-}0 \\
    1 && 0 && \phantom{-}0 && \phantom{-}0
  \end{pmatrix}\,, &
  \big(G_{\bar 1}\big)^{mn} &= \frac{1}{\sqrt{2}} \begin{pmatrix}
    \phantom{-}0 && 1 && 0 && \phantom{-}0 \\
    -1 && 0 && 0 && \phantom{-}0 \\
    \phantom{-}0 && 0 && 0 && -1 \\
    \phantom{-}0 && 0 && 1 && \phantom{-}0
  \end{pmatrix}\,, \nonumber \\
  \big(G_{\bar 2}\big)^{mn} &= \frac{1}{\sqrt{2}} \begin{pmatrix}
    \phantom{-}0 && \phantom{-}0 && 1 && 0 \\
    \phantom{-}0 && \phantom{-}0 && 0 && 1 \\
    -1 && \phantom{-}0 && 0 && 0 \\
    \phantom{-}0 && -1 && 0 && 0
  \end{pmatrix}\,, &
  \big(G_{\bar 3}\big)^{mn} &= \frac{1}{\sqrt{2}} \begin{pmatrix}
    \phantom{-}0 && 0 && \phantom{-}0 && 1 \\
    \phantom{-}0 && 0 && -1 && 0 \\
    \phantom{-}0 && 1 && \phantom{-}0 && 0 \\
    -1 && 0 && \phantom{-}0 && 0
  \end{pmatrix}
\end{align}
and satisfy the identities
\begin{align}
\big( G_{A} \big)_{mn} \big( G_{B} \big)^{mn} &= 2 \eta_{AB}\,, \\
\big( G_{A} \big)_{mp} \big( G_{B} \big)^{pn} + \big( G_{B} \big)_{mp} \big( G_{A} \big)^{pn} &= - \delta_m{}^n \, \eta_{AB}
\end{align}
with the standard O($D,D$) invariant metric
\begin{equation}
  \eta_{AB} = \begin{pmatrix}
    \delta_{ab} & 0 \\ 0 & -\delta_{\bar{a}\bar{b}}
  \end{pmatrix}
\end{equation}
of DFT${}_\mathrm{WZW}$. With them, we finally obtain the covariant fluxes
\begin{equation}\label{eqn:SL(4)toSO(3,3)}
  F_{ABC} = \left( X_{mn} \right)_{pq}{}^{rs}
    \left( G_A \right)^{mn} \left( G_B \right)^{pq} \left( G_C \right)_{rs}
\end{equation}
in their familiar form.

For our setup, the quadratic constraint of the embedding tensor is equivalent to the Jacobi identity \eqref{eqn:jacobiid} for the structure coefficients $F_{ABC}$ of the background vielbein. In the $\mathfrak{sl}(4)$ representation \eqref{eqn:embsl42} discussed above, the Jacobi identity has the simple form \cite{Dibitetto:2012rk}
\begin{equation}\label{eqn:embsl4jacobi}
  M_{mp} \tilde{M}^{pn} = \frac{1}{4} \delta_m{}^n M_{pq} \tilde{M}^{pq}\,.
\end{equation}
Since the matrix $M_{np}$ is symmetric, one can always find a SO($4$) rotation to diagonalize it. This group is the maximal subgroup of SL($4$) and is up to $\mathds{Z}_2$ isomorphic to SO($3$)$\times$SO($3$), the maximal compact subgroup of SO($3,3$). Hence, there is always a double Lorentz transformation that can be applied to the structure coefficients to diagonalize $M_{np}$. If $M_{np}$ is diagonal, $\tilde{M}^{rq}$ is diagonal, too. Otherwise, equation \eqref{eqn:embsl4jacobi} would be violated. This observation allows us to solve the quadratic constraint.
\begin{table}[t!]
\centering
\begin{tabular}{| c | r r r r | r r r r | c | c |}
\hline
\textrm{ID} & \multicolumn{4}{|c|}{$\diag M_{mn}/\,\cos\alpha\,$} & \multicolumn{4}{|c|}{$\diag \tilde{M}^{mn}/\,\sin\alpha\,$} & range of $\alpha$ & gauging \\[1mm]
\hline \hline \rowcolor{fillcolor}
$1$ & $1$ & $1$ & $1$ & $1$ & 
      $1$ & $1$ & $1$ & $1$ & $-\frac{\pi}{4}\,<\,\alpha\,\le\,\frac{\pi}{4}$ & $\left\{%
\setlength{\arraycolsep}{2pt}\begin{array}{cc}\textrm{SO}($4$)\ , & \alpha\,\ne\,\frac{\pi}{4}\ ,\\ \textrm{SO}(3)\ , & \alpha\,=\,\frac{\pi}{4}\ .\end{array}\right.$\\[4mm]
\hline
$2$ & $1$ & $1$ & $1$ & $-1$ & 
      $1$ & $1$ & $1$ & $-1$ & $-\frac{\pi}{4}\,<\,\alpha\,\le\,\frac{\pi}{4}$ & SO($3,1$)\\[1mm]
\hline
$3$ & $1$ & $1$ & $-1$ & $-1$ &
      $1$ & $1$ & $-1$ & $-1$ & $-\frac{\pi}{4}\,<\,\alpha\,\le\,\frac{\pi}{4}$ & $\left\{%
\setlength{\arraycolsep}{2pt}\begin{array}{cc}\textrm{SO}($2,2$)\ , & \alpha\,\ne\,\frac{\pi}{4}\ ,\\ \textrm{SO}(2,1)\ , & \alpha\,=\,\frac{\pi}{4}\ .\end{array}\right.$\\[2mm]
\hline \hline \rowcolor{fillcolor}
$4$ & $1$ & $1$ & $1$ & $0$ & 
      $0$ & $0$ & $0$ & $1$ & $-\frac{\pi}{2}\,<\,\alpha\,<\,\frac{\pi}{2}$ & ISO($3$)\\[1mm]
\hline
$5$ & $1$ & $1$ & $-1$ & $0$ & 
      $0$ & $0$ & $0$  & $1$ & $-\frac{\pi}{2}\,<\,\alpha\,<\,\frac{\pi}{2}$ & ISO($2,1$)\\[1mm]
\hline \hline \rowcolor{fillcolor}
$6$ & $1$ & $1$ & $0$ & $0$ & 
      $0$ & $0$ & $1$ & $1$ & $-\frac{\pi}{4}\,<\,\alpha\,\le\,\frac{\pi}{4}$ & $\left\{%
\setlength{\arraycolsep}{2pt}\begin{array}{cc}\textrm{CSO}(2,0,2)\ , & \alpha\,\ne\,\frac{\pi}{4}\ ,\\ \mathfrak{f}_{1}\quad(\textrm{Solv}_{6}) \ , & \alpha\,=\,\frac{\pi}{4}\ .\end{array}\right.$\\[4mm]
\hline
$7$ & $1$ & $1$ & $0$ & $0$ & 
      $0$ & $0$ & $1$,& $-1$ & $-\frac{\pi}{2}\,<\,\alpha\,<\,\frac{\pi}{2}$ & $\left\{%
\setlength{\arraycolsep}{2pt}\begin{array}{cc}\textrm{CSO}(2,0,2)\ , & |\alpha|\,<\,\frac{\pi}{4}\ ,\\ \textrm{CSO}(1,1,2)\ , & |\alpha|\,>\,\frac{\pi}{4}\ ,\\ \mathfrak{g}_{0}\quad(\textrm{Solv}_{6}) \ , & |\alpha|\,=\,\frac{\pi}{4}\ .\end{array}\right.$\\[4mm]
\hline \rowcolor{fillcolor}
$8$ & $1$ & $1$ & $0$ & $0$ & 
      $0$ & $0$ & $0$ & $1$ & $-\frac{\pi}{2}\,<\,\alpha\,<\,\frac{\pi}{2}$ & $\mathfrak{h}_{1}\quad(\textrm{Solv}_{6})$\\[1mm]
\hline
$9$ & $1$ & $-1$ & $0$ & $0$ & 
      $0$ & $0$  & $1$ & $-1$ & $-\frac{\pi}{4}\,<\,\alpha\,\le\,\frac{\pi}{4}$ & $\left\{%
\setlength{\arraycolsep}{2pt}\begin{array}{cc}\textrm{CSO}(1,1,2)\ , & \alpha\,\ne\,\frac{\pi}{4}\ ,\\ \mathfrak{f}_{2}\quad(\textrm{Solv}_{6}) \ , & \alpha\,=\,\frac{\pi}{4}\ .\end{array}\right.$\\[4mm]
\hline
$10$ & $1$ & $-1$ & $0$ & $0$ & 
       $0$ & $0$  & $0$ & $1$ & $-\frac{\pi}{2}\,<\,\alpha\,<\,\frac{\pi}{2}$ & $\mathfrak{h}_{2}\quad(\textrm{Solv}_{6})$\\[1mm]
\hline \hline \rowcolor{fillcolor}
$11$ & $1$ & $0$ & $0$ & $0$ & 
       $0$ & $0$ & $0$ & $1$ &
$-\frac{\pi}{4}\,<\,\alpha\,\le\,\frac{\pi}{4}$ &
$\left\{%
\setlength{\arraycolsep}{2pt}\begin{array}{cc}\mathfrak{l}\quad(\textrm{Nil}_{6}(3)\,)\ , & \alpha\,\ne\,0\ ,\\
\textrm{CSO}(1,0,3)\ , &
\alpha\,=\,0\ .\end{array}\right.$\\[4mm]
\hline \hline \rowcolor{fillcolor}
$12$ & $0$ & $0$ & $0$ & $0$ & 
       $0$ & $0$ & $0$ & $0$ &
$\alpha = 0$ & 
$\textrm{U}(1)^6$ \\
\hline
\end{tabular}
\caption{Solutions of the embedding tensor for half-maximal, electrically gauged supergravity in $n=3$ dimensions. All shaded entries give rise to compact groups. Details about $\mathfrak{f}_{1}$, $\mathfrak{f}_{2}$, $\mathfrak{g}_{0}$, $\mathfrak{h}_{1}$ and $\mathfrak{h}_{2}$ can be found in \cite{Dibitetto:2012rk}. All compact solution are also discussed in appendix~\ref{app:twists} in detail.}\label{tab:solembedding}
\end{table}
In total, one finds the eleven different non-trivial solutions \cite{Dibitetto:2012rk} presented in table~\ref{tab:solembedding}. All of them depend on one real parameter $\alpha$. The shaded ones are compact\footnote{Note that groups like ISO($3$) or CSO($2,0,2$) are of course in general not compact. However, one is able to make them compact by identifying various points. In the same way a compact $D$-tours arises from the non-compact plane $\mathds{R}^D$. As discussed e.g. in \cite{Hassler:2014sba}, this procedure puts restrictions on the background fluxes and quantizes them.} and thus the appropriate starting point for a compactification. For completeness, we also added the trivial solution $12$ with vanishing structure coefficients. It arises after a compactification on a $\mathrm{T}^3$. Note that only the solutions $1, 2$ and $3$ give rise to semisimple Lie groups. The others correspond to solvable and nilpotent Lie groups. Appendix~\ref{app:twists} shows how to construct the DFT${}_\mathrm{WZW}$ background generalized vielbein $E_A{}^I$ for all shaded, compact solutions. 

\subsection{Original DFT}\label{sec:traditionalSS}
In this subsection, we review generalized Scherk-Schwarz compactifications in the original flux formulation. In order to perform a compactification, it is essential to distinguish between internal, compact and external, extended directions. In the following we assume that there are $n$ internal and $D-n$ external ones. To make this situation manifest, we split the flat and curved doubled indices used in original DFT into the components
\begin{equation}
  V^{\bar A} = \begin{pmatrix} V_a & V^a & V^A \end{pmatrix}
    \quad \text{and} \quad
  W^{\bar I} = \begin{pmatrix} W_\mu & W^\mu & W^I \end{pmatrix}\,.
\end{equation}
Lowercase indices like $a$ and $\mu$ describe external directions and thus run from $0$ to $D-1$, while $A$ and $I$ parameterized the internal, $2n$-dimensional doubled space. In this convention, the O($D,D$) invariant metric reads
\begin{equation}
  \eta_{\bar M\bar N}=\begin{pmatrix}
    0 & \delta^\mu_\nu & 0 \\
    \delta_\mu^\nu & 0 & 0 \\
    0 & 0 & \eta_{MN} 
  \end{pmatrix}\,, \qquad
  \eta^{\bar M \bar N}=\begin{pmatrix}
    0 & \delta_\mu^\nu & 0\\
    \delta^\mu_\nu & 0 & 0\\
    0 & 0 & \eta^{MN} 
  \end{pmatrix}
\end{equation}
and the flat generalized metric is defined as
\begin{equation}
  S_{\bar A\bar B}=\begin{pmatrix}
    \eta^{ab} & 0 & 0 \\
    0 & \eta_{ab} & 0 \\
    0 & 0 & S_{AB}
  \end{pmatrix}\,, \qquad
  S^{\bar A\bar B}=\begin{pmatrix}
    \eta_{ab} & 0 & 0 \\
    0 & \eta^{ab} & 0 \\
    0 & 0 & S^{AB}
  \end{pmatrix}\,.
\end{equation} 
The curved version of the generalized metric arises after applying the twisted generalized vielbein \cite{Geissbuhler:2011mx,Aldazabal:2013sca,Geissbuhler:2013uka}
\begin{equation} \label{eqn:twistscherkschw}
  E^{\bar A}{}_{\bar M}(X) = 
    \widehat E^{\bar A}{}_{\bar N}(\mathds{X}) 
    U^{\hat N}{}_{\hat M}(\mathds{Y}) 
  \quad \text{with} \quad
  U^{\hat N}{}_{\hat M} = 
  \begin{pmatrix}
    \delta^\mu_\nu & 0 & 0 \\
      0 & \delta_\mu^\nu & 0 \\
      0 & 0 & U^N{}_M
  \end{pmatrix}
\end{equation}
to the flat version $S_{\bar A\bar B}$, resulting in
\begin{equation}
  \mathcal{H}_{\bar M\bar N} = E^{\bar A}{}_{\bar M} S_{\bar A\bar B} E^{\bar B}{}_{\bar N}\,.
\end{equation}
This twisted vielbein implements a special case of the generalized Kaluza-Klein ansatz \cite{Hohm:2013nja} called generalized Scherk-Schwarz ansatz. It is a product of two parts: While the generalized vielbein
\begin{equation}\label{eqn:vielbeinKKansatz}
  \widehat{E}^{\bar A}{}_{\bar M} = \begin{pmatrix}
    e_\alpha{}^\mu & - e_\alpha{}^\rho C_{\mu\rho}
      & - e_\alpha{}^\rho \widehat{A}_{M\rho} \\
    0 & e^\alpha{}_\mu & 0 \\
    0 & \widehat{E}^A{}_L \widehat{A}^L{}_\mu & 
    \widehat{E}^A{}_M
  \end{pmatrix} \quad \text{with} \quad C_{\mu\nu} = B_{\mu\nu} + 
  \frac{1}{2} \widehat{A}^L{}_\mu \widehat{A}_{L\nu} \,,
\end{equation}
which combines all dynamic fields of the effective theory, only depends on the external coordinates $\mathds{X}$, the twist $U^N{}_M$ just depends on the internal coordinates $\mathds{Y}$. All quantities it has a non-trivial action on are induced by a hat. For simplicity, we assume that the generalized dilaton $d$ is constant in the internal space. Moreover, the twist further has to fulfill the following constraints \cite{Geissbuhler:2011mx,Dibitetto:2012rk,Aldazabal:2011nj,Aldazabal:2013sca}:
\begin{itemize}
  \item Only O($n,n$)-valued twists with the defining property
    \begin{equation}\label{eqn:twistprop1}
      U_I{}^K \eta_{KL} U_J{}^L = \eta_{IJ}
    \end{equation}
    are allowed.
  \item The structure coefficients of the effective theory's gauge algebra
    \begin{equation}
      F_{IJK} = 3 U_{[I}{}^L \partial_L U_J{}^M U_{K]M} = \text{const.}
    \end{equation}
    have to be constant.
  \item The structure coefficients have to fulfill the Jacobi identity
    \begin{equation}\label{eqn:twistprop3}
      F_{M[IJ} F^M{}_{K]L} = 0\,.
    \end{equation}
\end{itemize}
Note that these properties imply that the structure coefficients $F_{IJK}$ are solutions of the embedding tensor, which we discussed in the last subsection.

Using them, one is able to calculate all components of the covariant fluxes
\begin{align}
  \mathcal{F}_{\bar A\bar B\bar C} &= 3 E_{[\bar A}{}^I \partial_I E_{\bar B}{}^J E_{\bar C]J} \quad \text{and} \\
  \mathcal{F}_{\bar A} &= E^{\bar BI}\partial_I E_{\bar B}{}^J E_{\bar A J} + 2 E_{\bar A}{}^I \partial_I d \quad \text{with} \quad d = \phi - \frac{1}{2} \log \det e^a{}_\mu\,.
\end{align}
Remember that these two definitions differ significantly from the ones used in DFT${}_\mathrm{WZW}$. After some algebra, one obtains the non-vanishing flux components\cite{Geissbuhler:2011mx,Hohm:2013nja}
\begin{align}
  \mathcal{F}_{abc} &= e_a{}^\mu e_b{}^\nu e_c{}^\rho \widehat{G}_{\mu\nu\rho} &
  \mathcal{F}_{ab}{}^c &= 2 e_{[a}{}^\mu \partial_\mu e_{b]}{}^\nu e^c{}_\nu = f_{ab}^c \nonumber \\
  \mathcal{F}_{abC} &= -e_a{}^\mu e_b{}^\nu \widehat{E}_{CM} \widehat{F}^M{}_{\mu\nu} &
  \mathcal{F}_{aBC} &= e_a{}^\mu \widehat{D}_\mu \widehat{E}_B{}^M \widehat{E}_{CM} \nonumber \\
  \mathcal{F}_{ABC} &= 3 \Omega_{[ABC]} &
  \mathcal{F}_{a}   &= f^{b}_{ab} + 2 e_a{}^\mu \partial_\mu \phi\,. \label{eqn:fluxcomponents}
\end{align}
These equations are written in a manifest gauge covariant way, by using the gauge covariant derivative
\begin{equation}
  \widehat{D}_\mu \widehat{E}_A{}^M = \partial_\mu \widehat{E}_A{}^M  - \mathcal{F}^M{}_{JI} \widehat{A}^J{}_\mu \widehat{E}_A{}^I\,.
\end{equation}
The corresponding field strength
\begin{equation}
  \widehat{F}^M{}_{\mu\nu} = 2 \partial_{[\mu}
  \widehat{A}^M{}_{\nu]} -
    \mathcal{F}^M{}_{NL} \widehat{A}^N{}_\mu
    \widehat{A}^L{}_\nu
\end{equation}
is defined as usual in Yang-Mills theories. It fulfills the Bianchi identity
\begin{equation}
  D_{[\mu} F^M{}_{\nu\rho]} = 0\,.
\end{equation}
Furthermore, the canonical field strength for the $B$-field, $B_{ij}$, is extended by a Chern-Simons term in order to be invariant under gauge transformations. The resulting 3-form
\begin{equation}
  \widehat{G}_{\mu\nu\rho} = 
  3\partial_{[\mu} B_{\nu\rho]} + 3\partial_{[\mu} 
    \widehat{A}^M{}_\nu
    \widehat{A}_{M \rho ]} - \mathcal{F}_{MNL} 
    \widehat{A}^M{}_\mu
    \widehat{A}^N{}_\nu
    \widehat{A}^L{}_\rho
\end{equation}
also fulfills a Bianchi identity, namely
\begin{equation}
  \partial_{[\mu} G_{\nu\rho\lambda]} = 0\,.
\end{equation}
Plugging the covariant flux \eqref{eqn:fluxcomponents} into the action of the original flux formulation
\begin{align}
  S = \int d^{2D} X\, e^{-2d} &\big( \mathcal{F}_A \mathcal{F}_B S^{AB} + \frac{1}{4} \mathcal{F}_{ACD} \mathcal{F}_B{}^{CD} S^{AB} - \frac{1}{12} \mathcal{F}_{ABC} \mathcal{F}_{DEF} S^{AD} S^{BE} S^{CF} \nonumber \\ \label{eqn:Sdftfluxform}
    & \quad - \frac{1}{6} \mathcal{F}_{ABC} \mathcal{F}^{ABC} - \mathcal{F}_A \mathcal{F}^A \big)
\end{align}
and switching to curved indices, we finally arrive at the effective action \cite{Geissbuhler:2011mx,Hohm:2013nja} 
\begin{align}\label{eqn:lowdimeffaction}
  S_\mathrm{eff} = \int d^{D-n}x \sqrt{-g} \, e^{-2\phi} \Big( &R + 4 \partial_\mu \phi \, \partial^\mu \phi - \frac{1}{12} \widehat{G}_{\mu\nu\rho} \widehat{G}^{\mu\nu\rho} \nonumber\\
  &-\frac{1}{4} \widehat{\mathcal H}_{MN} \widehat{\mathcal F}^{M \mu\nu} \widehat{\mathcal F}^N{}_{\mu\nu} + \frac{1}{8} \widehat{\mathcal D}_\mu \widehat{\mathcal H}_{MN} \widehat{\mathcal D}^\mu \widehat{\mathcal H}^{MN} - V \Big)\,,
\end{align}
with the scalar potential
\begin{equation}
  V = - \frac{1}{4} {F_I}^{KL} F_{JKL} \widehat{\mathcal H}^{IJ} + \frac{1}{12} F_{IKM} F_{JLN} \widehat{\mathcal H}^{IJ} \widehat{\mathcal H}^{KL} \widehat{\mathcal H}^{MN} + \frac{1}{6} F_{IJK} F^{IJK}\,.
\end{equation}
Here, $R$ denotes the standard scalar curvature in the external directions. As a consequence of the generalized Scherk-Schwarz ansatz, the Lagrange density of DFT is constant in the internal directions. Thus, it is trivial to solve the action's integral in these directions. The resulting global factor is neglected. As expected, the action \eqref{eqn:lowdimeffaction} describes a bosonic subsector of a half-maximal, electrically gauged supergravity. It is equivalent to the one presented by \cite{Aldazabal:2011nj}.

Note that all derivations in this subsection only took into account the properties \eqref{eqn:twistprop1}-\eqref{eqn:twistprop3} of the twist $U_I{}^J$. However, it is in general not clear whether twists with exactly these properties exist for all solutions of the embedding tensor. There is no systematic way to construct them. One is left with guessing solutions for the partial differential equation \eqref{eqn:twistprop1} which are elements of O($n,n$) at the same time. Some of these solutions were discussed in \cite{Dibitetto:2012rk,Hassler:2014sba} and more recently in the context of Extended Field Theory (EFT) \cite{Hohm:2014qga}. This problem concerning the twist is a major difference between geometric Scherk-Schwarz compactifications \cite{Scherk:1978ta,Scherk:1979zr}, which have been known for many years in the context of supergravity compactifications, and their generalization in DFT. For the former, there is a straightforward way to construct the twist. One uses the right or left invariant Maurer-Cartan form on the group manifold the compactification is performed on. Unfortunately, this procedure is not applicable to original DFT, because it requires a geometry ruled by ordinary diffeomorphisms and not by generalized diffeomorphisms. In the remainder of this paper, we will show that our new formulation cures this problem. Because all background fields transform covariantly under $2D$-diffeomorphism, we recover the common notion of geometry. Thus, as subsection~\ref{sec:twistconstr} shows, one is again able to use the right or left invariant Maurer Cartan form.

\subsection{DFT on group manifolds}\label{sec:genSSDFTWZW}
Equipped with the flux formulation of DFT${}_\mathrm{WZW}$, we now perform a generalized Scherk-Schwarz compactification and present the resulting low energy effective action. Throughout the following calculations, we have to distinguish between $n$ compact, internal directions and $D-n$ extended, external directions, corresponding to the internal coordinates $\mathds{Y}$ and external coordinates $\mathds{X}$, respectively. To make this situation manifest, we split the three different types of indices, which are relevant for the flux formulation derived in the last sections, according to
\begin{align}\label{eqn:genSSindexconv}
  V^{\hat{\tilde A}} &= \begin{pmatrix} V_{\hat a} & V^{\hat a} & V_{\hat A} \end{pmatrix} &
  W^{\tilde B} &= \begin{pmatrix} W_b & W^b & W^B \end{pmatrix} &
  X^{\tilde M} &= \begin{pmatrix} X_\mu & X^\mu & X^M \end{pmatrix}\,. \\
  \intertext{This step is equivalent to the strategy in DFT. There is only the difference that we have to treat three different kinds of indices (hatted, flat and curved) with this splitting, while DFT has only two, as it does not possess a background vielbein. The external indices $\hat a$, $a$ and $\mu$ run from $0$ to $D-n-1$ and their internal counterparts $\hat A$, $A$ and $M$ parameterize a $2n$-dimensional, doubled space. This index convention gives rise to three different versions of the $\eta$-metric}
  \eta_{\hat{\tilde A}\hat{\tilde B}} &= \begin{pmatrix} 0 & \delta^{\hat a}{}_{\hat b} & 0 \\  
    \delta_{\hat a}{}^{\hat b} & 0 & 0 \\ 0 & 0 & \eta_{\hat A\hat B} \end{pmatrix} &
  \eta_{\tilde A\tilde B} &= \begin{pmatrix} 0 & \delta^a{}_b & 0 \\ 
    \delta_a{}^b & 0 & 0 \\ 0 & 0 & \eta_{AB} \end{pmatrix} &
  \eta_{\tilde M\tilde N} &= \begin{pmatrix} 0 & \delta^\mu{}_\nu & 0 \\
    \delta_\mu{}^\nu & 0 & 0 \\ 0 & 0 & \eta_{MN} \end{pmatrix}
\end{align}
that are used to lower the indices defined in \eqref{eqn:genSSindexconv}. Moreover, we use the flat, background generalized metric
\begin{equation}
  S_{\hat{\tilde A}\hat{\tilde B}}=\begin{pmatrix}
    \eta^{\hat a\hat b} & 0 & 0 \\
    0 & \eta_{\hat a\hat b} & 0 \\
    0 & 0 & S_{\hat A\hat B}
  \end{pmatrix} \quad \text{and its inverse} \quad
  S^{\hat{\tilde A}\hat{\tilde B}}=\begin{pmatrix}
    \eta_{\hat a\hat b} & 0 & 0 \\
    0 & \eta^{\hat a\hat b} & 0 \\
    0 & 0 & S^{\hat A\hat B}
  \end{pmatrix}\,.
\end{equation}
For the next step, we specify the Scherk-Schwarz ansatz of the composite generalized vielbein
\begin{equation}\label{eqn:WZWSSansatz}
  \mathcal{E}_{\hat{\tilde A}}{}^{\tilde M} = \tilde E_{\hat{\tilde A}}{}^{\tilde B}(\mathbb{X}) \, E_{\tilde B}{}^{\tilde M}(\mathbb{Y})\,. \end{equation}
Its fluctuation part only depends on the external coordinates $\mathds{X}$, while the background part only depends on the internal ones $\mathds{Y}$. In comparison with the ansatz in \cite{Geissbuhler:2011mx,Aldazabal:2011nj,Aldazabal:2013sca,Hohm:2013nja}, the background generalized vielbein $E_{\tilde B}{}^{\tilde M}$ takes the role of the twist $U^{\hat N}{}_{\hat M}$. As opposed to the twist, it is not restricted to be O($D,D$) valued. This observation solves the problem of constructing an appropriate twist: There is always a straightforward way to construct $E_{\tilde B}{}^{\tilde M}$ as the left-invariant Maurer Cartan form on a group manifold. We went through this process for the example of $S^3$ with $H$-flux in \cite{Hassler:2014sba}.

For the fluctuation vielbein $\tilde{E}_{\hat{\tilde A}}{}^{\tilde B}$, the generalized Kaluza-Klein ansatz \cite{Geissbuhler:2011mx,Aldazabal:2013sca,Hohm:2013nja} is adapted to the index structure introduced above and gives rise to
\begin{equation}\label{eqn:KKansatzWZW}
  \tilde{E}_{\hat{\tilde A}}{}^{\tilde B}(\mathbb{X}) = \begin{pmatrix}
    {e_b}{}^{\hat a} & 0 & 0 \\ - {e_{\hat a}}^c C_{bc} & {e_{\hat a}}^b & -{e_{\hat a}}^c \widehat{A}^B{}_c \\ \widehat{E}_{\hat A}{}^C \widehat{A}_{C b} & 0 & \widehat{E}_{\hat A}{}^B
\end{pmatrix}
    \quad \text{with} \quad
    C_{ab} = B_{ab} + \frac{1}{2} \widehat{A}^D{}_a \widehat{A}_{D b}\,.
\end{equation}
In this ansatz, $B_{ab}$ denotes the two-form field appearing in the effective theory and
\begin{equation}
  \widehat{\mathcal H}^{CD} = \widehat{E}_{\hat A}{}^C S^{\hat A\hat B} \widehat{E}_{\hat B}{}^D
\end{equation}
represents $n^2$ independent scalar fields which form the moduli of the internal space. Analogous to the twist, the background vielbein has only non-trivial components in the internal space and reads
\begin{equation}
\label{eqn:scherkbackgroundvielbein}
  E_{\tilde B}{}^{\tilde M}(\mathbb{Y}) = \begin{pmatrix}
    \delta^b{}_\mu & 0 & 0 \\ 0 & \delta_b{}^\mu & 0 \\ 0 & 0 & E_B{}^M
  \end{pmatrix}\,.
\end{equation}
With the Kaluza-Klein ansatz \eqref{eqn:KKansatzWZW} and the partial derivative
\begin{equation}
  \partial^{\tilde M} = \begin{pmatrix} \partial_\mu & \partial^\mu & \partial^M \end{pmatrix}\,
\end{equation}
in mind, it is straightforward to calculate the fluxes $\tilde{F}_{\hat{\tilde A}\hat{\tilde B}\hat{\tilde C}}$ and $\tilde{F}_{\hat{\tilde A}}$ defined in \eqref{eqn:fluxfluxes} and \eqref{eqn:defFA}. After some algebra, we obtain the non-vanishing components
\begin{align}
  \tilde{F}_{\hat a\hat b\hat c} &= e_{\hat a}{}^{d} e_{\hat b}{}^{e} e_{\hat c}{}^{f} \, 3 \big( D_{[d} B_{e f]} + \widehat{A}^D{}_{[d} D_e \widehat{A}_{D f]} \big) & 
  \tilde{F}_{\hat a\hat b}{}^{\hat c} &= 2 e_{[\hat a}{}^d D_{d} e_{\hat b]}{}^e e_{e}{}^{\hat c} = \tilde{f}_{\hat a\hat b}^{\hat c} \nonumber \\ 
  \tilde{F}_{\hat a\hat b\hat C} &= - e_{\hat a}{}^d e_{\hat b}{}^e \widehat{E}_{\hat CD} \, 2 D_{[d} \widehat{A}^D{}_{e]} &
  \tilde{F}_{\hat a\hat B\hat C} &=  e_{\hat a}{}^d D_d \widehat{E}_{\hat B}{}^D \widehat{E}_{\hat CD} \nonumber \\
  \tilde{F}_{\hat a} &= \tilde f_{\hat a\hat c}^{\hat c} + 2 e_{\hat a}{}^b D_b \phi\,. \label{eqn:genSSfluxfluxes}
\end{align}
In order to determine the covariant fluxes $\mathcal{F}_{\hat{\tilde A}\hat{\tilde B}\hat{\tilde C}}$, we also have to evaluate the background contribution $F_{\hat{\tilde A}\hat{\tilde B}\hat{\tilde C}}$. As the background vielbein \eqref{eqn:scherkbackgroundvielbein} only depends on internal coordinates, the only non-vanishing components of $F_{\tilde A\tilde B\tilde C}$ are
\begin{equation}
  F_{ABC} = 2 \Omega_{[AB]C}\,.
\end{equation}
They give rise to the non-vanishing components
\begin{align}
  F_{\hat a\hat b\hat c} &= -e_{\hat a}{}^{d} e_{\hat b}{}^{e} e_{\hat c}{}^{f} \widehat{A}_d{}^D \widehat{A}_e{}^E \widehat{A}_f{}^F F_{DEF} &
  F_{\hat a\hat b\hat C}  &= e_{\hat a}{}^{d} e_{\hat b}{}^{e} \widehat{A}_c{}^D \widehat{A}_d{}^E \widehat{E}_{\hat C}{}^F F_{DEF} \nonumber \\ 
  F_{\hat a\hat B\hat C}  &= -e_{\hat a}{}^{b} \widehat{A}_b{}^D \widehat{E}_{\hat B}{}^E \widehat{E}_{\hat C}{}^F F_{DEF} &
  F_{\hat A\hat B\hat C}  &= E_{\hat A}{}^D E_{\hat B}{}^E E_{\hat C}{}^F F_{DEF} \,.
\end{align}
Combining these results with \eqref{eqn:genSSfluxfluxes} and remembering the gauge covariant quantities
\begin{align}
  \widehat{D}_\mu \widehat{E}_{\hat A}{}^B &= \partial_\mu \widehat{E}_{\hat A}{}^B - F^{B}{}_{CD} \widehat{A}_\mu{}^C \widehat{E}_{\hat A}{}^D \nonumber \\ 
  \widehat{F}^A{}_{\mu\nu} &= 2\partial_{[\mu} \widehat{A}_{\nu]}{}^A - {F^A}_{BC} \widehat{A}_\mu{}^B \widehat{A}_\nu{}^C \nonumber \\ 
  \widehat{G}_{\mu\nu\rho} &= 3 \partial_{[\mu} B_{\nu\rho]} + \widehat{A}_{[\mu}{}^A \partial_\nu \widehat{A}_{\rho]A} - F_{ABC} \widehat{A}_\mu{}^A \widehat{A}_\nu{}^B \widehat{A}_\rho{}^C\,,
\end{align}
discussed in section~\ref{sec:traditionalSS}, we finally obtain
\begin{align}
  \mathcal{F}_{\hat a\hat b\hat c} &= e_{\hat a}{}^{\mu} e_{\hat b}{}^{\nu} e_{\hat c}{}^{\rho} \widehat{G}_{\mu\nu\rho} &
  \mathcal{F}_{\hat a\hat b}{}^{\hat c} &= 2 e_{[\hat a}{}^\mu \partial_{\mu} e_{\hat b]}{}^\nu e_{\nu}{}^{\hat c} \nonumber \\ 
  \mathcal{F}_{\hat a\hat b\hat C} &= -e_{\hat a}{}^\mu e_{\hat b}{}^\nu \widehat{E}_{\hat CA} \widehat{F}^A{}_{\mu\nu} &
  \mathcal{F}_{\hat a\hat B\hat C} &= e_{\hat a}{}^\mu \widehat{D}_\mu \widehat{E}_{\hat B}{}^A \widehat{E}_{\hat CA} \nonumber \\
  \mathcal{F}_{\hat A\hat B\hat C} &= \widehat{E}_{\hat A}{}^D \widehat{E}_{\hat B}{}^E \widehat{E}_{\hat C}{}^F \, F_{DEF} &
  \mathcal{F}_{\hat a} &= \tilde f_{ab}^b + 2 e_{\hat a}{}^\mu \partial_\mu \phi \label{eqn:fluxcompWZW}\,.
\end{align}
Note that the gauge covariant objects here carry indices $A,B,C,\cdots$ instead of $I,J,K,\cdots$, as they do in the last subsection. This is because they have to carry O($n,n$) indices, which are the former for DFT${}_\mathrm{WZW}$ (depicted in \eqref{eqn:groupindices}) and the latter in the original formulation. From this point on, all further calculations proceed as explained in the last subsection. Thus, when substituting our results into the flux formulation's action of DFT${}_\mathrm{WZW}$ \eqref{eqn:actionfluxform}, we again obtain the effective action \eqref{eqn:lowdimeffactionwzw}. As explained above, this time the indices $I,J,K,\cdots$ are substituted by $A,B,C,\cdots$. However, this difference is mere convention. Furthermore, the scalar potential
\begin{equation}
  V = -\frac{1}{4} F_A{}^{CD} F_{BCD} \widehat{\mathcal H}^{AB} + \frac{1}{2} F_{ACE} F_{BDF} \widehat{\mathcal H}^{AB} \widehat{\mathcal H}^{CD} \widehat{\mathcal H}^{EF}
\end{equation}
lacks the strong constraint violating term $1/6 F_{ABC} F^{ABC}$, which appears as a cosmological constant in gauged supergravities, even if we do not impose the strong constraint on the background field. In section~\ref{sec:missingFABCFABCterm}, we argued why this term is missing in our formulation. Anyhow, from a bottom up perspective it is totally legitimate to add it by hand to the action in the same way as it was done in the original flux formulation. It is perfectly compatible with all the symmetries of the theory.

Our new approach solves an ambiguity of generalized Scherk-Schwarz compactifications: In the DFT${}_\mathrm{WZW}$ framework, the twist is equivalent to the background generalized vielbein $E_A{}^I$. It is constructed in the same way as for conventional Scherk-Schwarz compactifications. This is possible, because the theory possesses standard $2D$-diffeomorphisms. Thus, all mathematical tools available for group manifolds are applicable. We immediately lose these tools, if we return to the original DFT formulation, because the extended strong constraint, necessary for this transition, breaks $2D$-diffeomorphism invariance. Hence, one is left with the problems outlined in subsection~\ref{sec:traditionalSS}.

All derivations performed so far in DFT${}_\mathrm{WZW}$ are top down. It started from full bosonic CSFT in \cite{Blumenhagen:2014gva,Blumenhagen:2015zma} and was reduced step by step until we finally arrived at the low energy effective action \eqref{eqn:lowdimeffactionwzw}. Thus, one is able to explicitly check the uplift of solutions of its equations of motion to full string theory. In doing so, we have to keep in mind that all results obtained so far are only valid at tree level. Consistency at loop level, e.g. a modular invariant partition function, gives rise to additional restrictions. There is another lesson which can be learned from the CFT side: We know that the background fluxes $F_{ABC}$ scale with $1/\sqrt{k}$, where $k$ denotes the level of the Ka\v{c}-Moody algebra on the world sheet. To make this property manifest, we decompose them into
\begin{equation}
  F_{ABC} = \frac{1}{\sqrt{k}} f_{ABC}
\end{equation}
and assume that the structure coefficients $f_{ABC}$ are normalized, e.g.
\begin{equation}
  f_{AC}{}^D f_{BD}{}^C = \frac{1}{2} \delta_{AB}\,.
\end{equation}
Now, the gauge covariant derivative reads
\begin{equation}
  \widehat{D}_\mu V^A = \partial_\mu V^A - \frac{1}{\sqrt{k}}f^{A}{}_{BC} \widehat{A}_\mu{}^B C^C\,.
\end{equation}
From this equation, we immediately read off the Yang-Mills coupling constant 
\begin{equation}
  g_\mathrm{YM} = \frac{1}{\sqrt{k}}\,.
\end{equation}
Remember, the geometric interpretation of DFT${}_\mathrm{WZW}$ only holds in the large level limit $k\gg 1$. The corresponding effective theory is weakly coupled and thus can be treated perturbatively. However, freezing out all fluctuations in the internal directions, which is exactly the case for generalized Scherk-Schwarz compactifications, our results extend to $k=1$. In this case, one has to reduce the number of external directions to cancel the total central charges of the bosons and the ghost system.

\subsection{Constructing the twist}\label{sec:twistconstr}
A major advantage of generalized Scherk-Schwarz compactifications in DFT${}_\mathrm{WZW}$ is the existence of a straightforward procedure to construct the background vielbein $E_A{}^I$, which replaces the twist in the original scenario, by starting from a solution of the embedding tensor. In the following, we present this scheme in detail.

Let us first assume that $t_A$ denotes $2n$ different $N\times N$ matrices which give rise to the algebra
\begin{equation}\label{eqn:genalgebra}
  [t_A, t_B] = t_A t_B - t_B t_C = F_{AB}{}^C t_C\,.
\end{equation}
Its structure coefficients are equivalent to an arbitrary solution of the embedding tensor \eqref{eqn:embeddingtensor}. In this case
\begin{equation}
  F_{AB}{}^D\eta_{DC} + F_{AC}{}^D\eta_{BD} = 0
\end{equation}
has to hold. Furthermore, we define a non-degenerate, bilinear, symmetric two-form
\begin{equation}
  \mathcal{K}(t_A, t_B) = \eta_{AB}
\end{equation}
on the vector space spanned by the matrices $t_A$. Later, we will explain how they and the two-form are realized. At the moment, these three definitions are sufficient. With them, it is evident that the background fluxes $F_{ABC}$ are given by
\begin{equation}
  F_{ABC} = \mathcal{K}(t_A, [t_B, t_C])\,.
\end{equation}

The second ingredient, required to derive the background generalized vielbein, is a group element $g\in G$ of the group $G$ representing the background. Therefore, we use the exponential map
\begin{equation}\label{eqn:expmap}
  g = \exp( t_A X^A ) = \sum\limits_{m=0}^\infty \frac{1}{m !} (t_A X^A)^n
\end{equation}
in order to derive it from the generators $t_A$. For compact groups this map is surjective onto the identity component $G_0$ of $G$. We assume that all groups we treat here are path-connect and thus $G_0$ and $G$ are equivalent. The map \eqref{eqn:expmap} becomes bijective, if we restrict the domain of the coordinates $X^I$ accordingly. In this case, each group element is labeled by a unique point in the coordinate space. The left invariant Maurer-Cartan form is defined as
\begin{equation}\label{eqn:leftinvmc}
  E_{A I} = \mathcal{K}(t_A, g^{-1} \partial_I g)\,.
\end{equation}
Using it to calculate
\begin{equation}
  \Omega_{ABC} = E_A{}^I \partial_I E_B{}^J E_{C J}\,,
\end{equation}
we obtain
\begin{align}
  \Omega_{ABC} &= -E_A{}^I \big[ \mathcal{K}(t_C, \partial_I g^{-1}\, g g^{-1}\, \partial_J g) + \mathcal{K}(t_C, g^{-1} \partial_I \partial_J g) \big] E_B{}^J \nonumber \\
  &= E_B{}^J \mathcal{K}( g^{-1} \partial_J g, \, t_C \, g^{-1}\partial_I g) E_A{}^I - \mathcal{K}(t_C, g^{-1}\partial_I\partial_J g) E_A{}^I E_B{}^J \nonumber \\
  &= \mathcal{K}(t_B, t_C t_A) - \mathcal{K}(t_C, g^{-1} \partial_I \partial_J g) E_A{}^I E_B{}^J\,.
\end{align}
The coefficients of anholonomy give rise to the correct background covariant fluxes, namely
\begin{equation}
  F_{ABC} = 2\Omega_{[AB]C} = \mathcal{K}(t_A, [t_B, t_C])\,.
\end{equation}
Thus, we indeed recover the correct identity for the background generalized vielbein $E_A{}^I$, with the left invariant Maurer-Cartan form \eqref{eqn:leftinvmc}.

As already stated, the generators $t_A$ of the Lie algebra are $N\times N$ matrices
\begin{equation}
  t_A = \begin{pmatrix}
   (t_A)_{11} & \cdots & (t_A)_{1N} \\
   \vdots & & \vdots \\
   (t_A)_{N1} & \cdots & (t_A)_{NN}
  \end{pmatrix}\,.
\end{equation}
In order to evaluate $\mathcal{K}(x, y)$ for arbitrary algebra elements $x, y \in \mathfrak{g}$, we need to expand them in terms of the generators, e.g.
\begin{equation}
  x = \sum\limits_{A=1}^{2 n} c_A t_A\,,
\end{equation}
where $c_A$ denotes the $2n$ expansion coefficients. It is convenient to rearrange the matrix $x$ into the vector
\begin{equation}
  x = \begin{pmatrix} 
    x_{11} & \cdots & x_{1N} & x_{2N} & \cdots & x_{NN}
  \end{pmatrix}
\end{equation}
and solve the linear system of equations
\begin{equation}
  c M = x \quad \text{with} \quad
  M = \begin{pmatrix}
    (t_1)_{11} & \cdots & (t_1)_{1N} & (t_1)_{2N} & \cdots & (t_1)_{NN} \\
    \vdots & & \vdots & \vdots & & \vdots \\
    (t_{2n})_{11} & \cdots & (t_{2n})_{1N} & (t_{2n})_{2N} & \cdots & (t_{2n})_{NN}
  \end{pmatrix} \quad \text{and} \quad 
  c = \begin{pmatrix} c_1 & \cdots c_{2n} \end{pmatrix}
\end{equation}
to calculate these coefficients. We are interested in a unique solution, thus the $2 n \times N^2$ matrix $M$ has to have full rank
\begin{equation}\label{eqn:fullrank}
  \rank M = 2n\,.
\end{equation}
Besides $\eqref{eqn:genalgebra}$, this equation gives a second constraint on the generators $t_A$. According to Ado's theorem \cite{Ado1949}, both can be satisfied for a finite $N$. Such representations are called faithful. We show, how one obtains them for semisimple and solvable Lie algebras in the next subsections, which are partly based on \cite{Hofer2012}. Appendix~\ref{app:twists} applies these techniques to all the compact solutions of the embedding tensor presented in table~\ref{tab:solembedding}.

\subsubsection{Semisimple algebras}\label{sec:semisimple}
For semisimple Lie algebras, the generators
\begin{equation}\label{eqn:gensemisimple}
  (t_A)_{BC} = F_{AB}{}^C
\end{equation}
can be read off directly from the structure coefficients. Doing so, we obtain the adjoint representation
\begin{equation}
  \adj_x y = [x,y] \quad \text{with} \quad x,\,y \in \mathfrak{g}
\end{equation}
in the basis spanned by all abstract generators. It has dimension $N=2n$ and is faithful if the center of the Lie algebra
\begin{equation}\label{eqn:center}
  Z(\mathfrak{g}) = \{x\in \mathfrak{g} \,|\, [x,y]=0 \,\, \text{for all} \,\, y\in \mathfrak{g}\} 
\end{equation}
is trivial. This is the case for semisimple Lie algebras. However, there are also non-semisimple ones, such as ISO($3$) which is discussed in appendix~\ref{app:ISO(3)}, with vanishing center. The matrix realization of their generators is given by \eqref{eqn:gensemisimple}, too.

In general, the adjoint representation is not the lowest dimensional one. E.g. for SO($4$), which we present in appendix~\ref{app:SO4}, the adjoint has $N=6$, while the fundamental representation is only $4$-dimensional. In the end, each of them works for our purpose. However, taking the smallest one simplifies the calculations considerably.

\subsubsection{Nilpotent Lie algebras}\label{sec:nilpotent}
For nilpotent Lie algebras, \eqref{eqn:gensemisimple} gives rise to generators $t_A$ which are not linear independent from each other. Thus, they are not faithful and violate \eqref{eqn:fullrank}. Before discussing how to obtain proper generators, let us first give a criterion to identify these algebras. To this end, consider the lower central series
\begin{equation}
  L_{m+1} = [ \mathfrak{g}, L_m ] \quad \text{with} \quad L_0 = \mathfrak{g}\,.
\end{equation}
It gives rise to the series
\begin{equation}\label{eqn:lowersubalg}
  \mathfrak{g} = L_0 \supseteq L_1 \supseteq L_2 \supseteq \dots
\end{equation}
of subalgebras. If this series terminates at a finite $k$ with $L_k=\{0\}$, the algebra $\mathfrak{g}$ is nilpotent of order $k$.

In the following we make use of the infinite dimensional universal enveloping algebra $U(\mathfrak{g})$ of the nilpotent Lie algebra $\mathfrak{g}$. According to the Poincar\'e-Birkhoff-Witt theorem, it is spanned by the ordered monomials
\begin{equation}
  t(\alpha) = t_1^{\alpha_1} t_2^{\alpha_2} \dots t_{2n}^{\alpha_{2n}} \quad \text{with} \quad \alpha \in \mathds{Z}^3_+\,.
\end{equation}
Via left multiplication
\begin{equation}\label{eqn:phi_x}
  \phi_x: U(\mathfrak{g}) \rightarrow U(\mathfrak{g}) \,, \quad 
    \phi_x (y) = x y \quad\text{with}\quad x \in \mathfrak{g}\,,
\end{equation}
algebra elements $x$ act faithful on the universal enveloping algebra. Ado's theorem states that even on the finite dimensional subspace
\begin{equation}\label{eqn:nilpotsubspaceV}
  V^k = \{ t(\alpha) \in U(\mathfrak{g}) \,|\, \ord t(\alpha) \le k \}
\end{equation}
of $U(\mathfrak{g})$, $\phi_x$ acts still faithful. Here one uses the order function
\begin{equation}
  \ord t(\alpha) = \sum\limits_m^{2n} \alpha_m \ord t_m\,, \quad
  \ord t_m = \max \{s \,|\, t_m \in L_{s-1} \}
    \quad \text{and} \quad
  \ord 1 = 0
\end{equation}
to fix this subspace. To finally obtain a $N=\dim V$-dimensional, faithful matrix representation of the generators $t_A$, we express the linear operator $\phi_{t_A}$ in the basis which spans $V^k$.

\subsubsection{Solvable Lie algebras}\label{sec:solvable}
Techniques from both, semisimple and nilpotent Lie algebras, find their application in the case of solvable Lie algebras, which are characterized by a derived series
\begin{equation}
  L^{m+1} = [L^m, L^m] \quad \text{with} \quad L^0 = \mathfrak{g}
\end{equation}
which terminates at a finite $k$ with $L^k=\{0\}$. Like \eqref{eqn:lowersubalg}, it gives rise to the series
\begin{equation}
  \mathfrak{g} = L^0 \supset L^1 \supset \cdots \supset L^{k-1} \supset \{0\}\,.
\end{equation}
of subalgebras. The first of them, $\mathfrak{n}=L^1$ is nilpotent. Thus, we expand the map \eqref{eqn:phi_x} for all its generators $t\in\mathfrak{n}$ in the basis \eqref{eqn:nilpotsubspaceV} to obtain their matrix representation. Furthermore, the adjoint representations $\adj_x = [x,y]$,
\begin{equation}
  \adj_x 1 = 0 \quad \text{and} \quad \adj_x y_1 \dots y_l = \sum\limits_{m=1}^l y_1 \dots y_{m-1}[x,y_m] y_{m+1} \dots y_l \quad \text{with} \quad
  y_m \in \mathfrak{n}
\end{equation}
of the remaining generators $x \ni \mathfrak{q}=\mathfrak{g}/\mathfrak{n}$ act faithful on $V^k$, too. Besides $\phi_t$, we also express $\adj_u$, $u\in\mathfrak{q}$, in the basis $V^k$ to complete the $N=\dim V$ dimensional matrix representation of the algebra. Note that all nilpotent Lie algebras are automatically solvable with $\mathfrak{q}=\{\}$.

\section{Conclusion and Outlook}\label{sec:conclusion}
During the course of this paper, we derived the flux formulation of DFT${}_\mathrm{WZW}$ and applied it to examine generalized Scherk-Schwarz compactifications. In contrast to the original flux formulation, we obtained new covariant fluxes. They split into a fluctuation part, which has to fulfill the strong constraint, and a background part, for which the Jacobi identity is sufficient. Furthermore, the covariant fluxes transform as scalars under generalized diffeomorphisms and $2D$-diffeomorphisms. The latter ones are missing completely in the original DFT framework. This result underpins the general structure of our theory. Starting point is a geometric, $2D$-dimensional, pseudo Riemannian manifold with split signature. This space is isomorphic to a Lie group which admits an embedding into the group O($D,D$). Its metric, $\eta_{IJ}$, reduces the structure group of the manifold from GL($2D$) to O($D,D$). All dynamic fields, such as the generalized metric, are build on this reduced structure and reduce it even further.

Original DFT, either the generalized metric or the flux formulation, is lacking this geometric interpretation of $\eta_{IJ}$. It starts directly with the fixed O($D,D$) structure. As a consequence, problems arise in the construction of the twist for generalized Scherk-Schwarz compactifications. We discussed these problems in detail and showed that they are solved naturally in the framework of DFT${}_\mathrm{WZW}$. Here, the background generalized vielbein takes the role of the twist. Due to the geometric structure of the background, it can be identified with the left invariant Maurer-Cartan form on the group manifold, used in the compactification. This observation is in perfect agreement with geometric Scherk-Schwarz reductions \cite{Scherk:1978ta,Scherk:1979zr}, which also use the left or right invariant Maurer-Cartan form. Moreover, only the Jacobi identify has to hold for the background. On a group manifold, it is equivalent to the closure constraint introduced in \cite{Grana:2012rr,Geissbuhler:2013uka,Aldazabal:2013sca}. Thus, embeddings into the full O($D,D$) group are accessible.

As a top down approach, DFT${}_\mathrm{WZW}$ was constructed in \cite{Blumenhagen:2014gva,Blumenhagen:2015zma} from CSFT. Thus, one is able to identify all fields in the theory with quantities on the world sheet of closed string theory. Doing so, e.g. allowed us to explain why the strong constraint violating term $1/6 F_{ABC} F^{ABC}$ vanishes in the action. Even more, this relation can be used to uplift non-geometric backgrounds. However, this uplift still only takes tree-level computations into account. Modular invariance of the torus partition function introduces additional constraints. Still, the connection between non-geometric fluxes and the structure coefficients of the Ka\v{c}-Moody algebra in the world sheet CFT, which were already suggested in \cite{Blumenhagen:2014gva}, is now evident.

In generalized Scherk-Schwarz compactifications, there are no fluctuations in the internal space $\mathds{Y}$. Hence, the generalized metric $\mathcal{H}_{AB}$ and the generalized dilaton $d$ do not depend on $\mathds{Y}$. In this case, the strong constraint is solved trivially. However, it seems that especially non-trivial solutions of the strong constraint, which differs significantly from the one in the original formulation, expose the full power of DFT${}_\mathrm{WZW}$. Studying them presumably would also lead to a better understanding as how dualities are implemented in this theory.

\acknowledgments
We would like to thank
Enrico Brehm,
Ralph Blumenhagen,
Daniel Junghans and
Christoph Mayrhofer
for helpful discussions. F.H. would like to thank the City University of New York and the Columbia University for hospitality during parts of this project. This work was partially supported by the ERC Advanced Grant ``Strings and Gravity'' (Grant.No. 32004), by TRR 33 "The Dark
Universe", by the DFG cluster of excellence ``Origin and Structure of the Universe'' and by the NSF Grant PHY-1452037.

\appendix

\section{Twists for the compact solutions of the \texorpdfstring{O($3,3$)}{O(3,3)} embedding tensor}\label{app:twists}
In the following, we apply the techniques presented in section~\ref{sec:twistconstr} to derive the background generalized vielbeins $E_A{}^I$ for all compact solutions of the $n=3$ embedding tensor in table~\ref{tab:solembedding}. To this end, we first calculate the structure coefficients $F_{ABC}$ using \eqref{eqn:SL(4)toSO(3,3)}. In most cases, it is convenient to further apply a particular O($3,3$) rotation $R_A{}^B$ in order to simplify the results:
\begin{equation}
  F'_{ABC} = R_A{}^D R_B{}^E R_C{}^F F_{DEF} \quad \text{and} \quad
  \eta'_{AB} = R_A{}^D R_B{}^E \eta_{DE}\,.
\end{equation}
Moreover, we assign symbols to all six generators, e.g.
\begin{equation}
  t_A = \{a, b, c, d, e, f\}\,,
\end{equation}
and read off their algebra according to \eqref{eqn:genalgebra}. Starting from this algebra, we derive a $N$-dim\-ensio\-nal matrix representation for the generators by following the procedures outlined in sections~\ref{sec:semisimple}-\ref{sec:solvable}. Next, we obtain the group elements $g$ by applying the exponential map \eqref{eqn:expmap} and finally use them to calculate the left invariant Maurer-Cartan from \eqref{eqn:leftinvmc}.

\subsection{\texorpdfstring{SO($4$)}{SO(4)}/\texorpdfstring{SO($3$)}{SO(3)}}\label{app:SO4}
Applying the rotation
\begin{equation}
  R_A{}^B = \begin{pmatrix}
    \phantom{-}0 & & \phantom{-}0 & & \phantom{-}0 & & \phantom{-}0 & & \phantom{-}0 & & \phantom{-}1 \\
    \phantom{-}0 & & \phantom{-}0 & & \phantom{-}0 & & \phantom{-}0 & & \phantom{-}1 & & \phantom{-}0 \\
    \phantom{-}0 & & \phantom{-}0 & & \phantom{-}0 & & \phantom{-}1 & & \phantom{-}0 & & \phantom{-}0 \\
    \phantom{-}0 & & \phantom{-}0 & & -1 & & \phantom{-}0 & & \phantom{-}0 & & \phantom{-}0 \\
    \phantom{-}0 & & -1 & & \phantom{-}0 & & \phantom{-}0 & & \phantom{-}0 & & \phantom{-}0 \\
    -1 & & \phantom{-}0 & & \phantom{-}0 & & \phantom{-}0 & & \phantom{-}0 & & \phantom{-}0
\end{pmatrix}\,, \quad \text{results in} \quad
  \eta'_{AB} = \begin{pmatrix} 
    -1 & \phantom{-}0 & \phantom{-}0 & \phantom{-}0 & \phantom{-}0 & \phantom{-}0\\
     \phantom{-}0 & -1 & \phantom{-}0 & \phantom{-}0 & \phantom{-}0 & \phantom{-}0\\
     \phantom{-}0 & \phantom{-}0 & -1 & \phantom{-}0 & \phantom{-}0 & \phantom{-}0\\
     \phantom{-}0 & \phantom{-}0 &  \phantom{-}0 & \phantom{-}1 & \phantom{-}0 & \phantom{-}0 \\
     \phantom{-}0 & \phantom{-}0 &  \phantom{-}0 & \phantom{-}0 & \phantom{-}1 & \phantom{-}0 \\
     \phantom{-}0 & \phantom{-}0 &  \phantom{-}0 & \phantom{-}0 & \phantom{-}0 & \phantom{-}1
  \end{pmatrix}
\end{equation}
and the semisimple Lie algebra
\begin{equation}
\label{eqn:so4generators}
  [ s_a ,s_b ] = \alpha_-\, \varepsilon_{ab}{}^c \, s_c 
    \quad \text{and} \quad
  [ \bar s_a ,\bar s_b ] = \alpha_+\,\varepsilon_{ab}{}^c \, \bar s_c
\end{equation}
with
\begin{equation}
  \alpha_+ = - \sqrt{2} \big( \cos(\alpha) + \sin(\alpha) \big)
    \quad \text{and} \quad
  \alpha_- =  \sqrt{2} \big( \cos(\alpha) - \sin(\alpha) \big)\,,
\end{equation}
after assigning the symbols
\begin{equation}
  t_A = \{ s_1, s_2, s_3, \bar s_1, \bar s_2, \bar s_3 \}
\end{equation}
for the generators. Here $\varepsilon_{ab}{}^c$ denotes the totally antisymmetric tensor in three dimensions with $\varepsilon_{12}{}^3=1$. For $\alpha = 0$, this Lie algebra is equivalent to $\mathfrak{so}(4)$. Further, it degenerates at $\alpha = \pi / 4$ to $\mathfrak{so}(3)$. In the basis we have chosen, the decomposition
\begin{equation}
  \mathfrak{so}(4)_\alpha = \mathfrak{so}(3)_{\alpha_+} \oplus \mathfrak{so}(3)_{\alpha_-}
\end{equation}
is manifest.

We use the adjoint representation of the Lie algebra to construct the group elements. But instead of applying the exponential map \eqref{eqn:expmap}, we use
\begin{equation}\label{eqn:prodexpmap}
  g =  \exp(t_6 X^6) \exp(t_5 X^5) \cdots \exp(t_1 X^1)\,,
\end{equation}
which allows to read off the inverse group element
\begin{equation}
  g^{-1} = \exp(-t_1 X^1) \exp(-t_2 X^2) \cdots \exp(-t_6 X^6)
\end{equation}
directly. The coordinates
\begin{equation}
  X^I = \{ \phi_1, \phi_2, \phi_3, \bar\phi_1, \bar\phi_2, \bar\phi_3 \}\,.
\end{equation}
split into three angles twice, describing a rotation in $\mathds{R}^3$ each. Finally, we construct the left invariant Maurer-Cartan form
\begin{equation}
  E^A{}_I = \begin{pmatrix}
    A_{\alpha-}(\phi_1, \phi_2, \phi_3) & 0 \\ 0 & A_{\alpha+}(\bar\phi_1, \bar\phi_2, \bar\phi_3)
\end{pmatrix}
\end{equation}
and its inverse transposed
\begin{equation}
  E_A{}^I = \begin{pmatrix}
    A_{\alpha-}^{-T}(\phi_1, \phi_2, \phi_3) & 0 \\ 0 & A_{\alpha+}^{-T}(\bar\phi_1, \bar\phi_2, \bar\phi_3)
  \end{pmatrix}
\end{equation}
where $A_\alpha$ denotes the matrix
\begin{equation}\label{eqn:Aalpha}
A_\alpha(\phi_1,\phi_2,\phi_3) = \begin{pmatrix}
    \phantom{-}1 & 0 & -\sin(\phi_2\alpha) \\
    0 & \phantom{-}\cos(\phi_1\alpha) & \phantom{-}\cos(\phi_2\alpha) \sin(\phi_1\alpha) \\
    0 & -\sin(\phi_1\alpha) & \phantom{-}\cos(\phi_1\alpha) \cos(\phi_2\alpha)
  \end{pmatrix}
\end{equation}
and its inverse transpose reads
\begin{equation}
A_\alpha^{-T}(\phi_1,\phi_2,\phi_3) = \begin{pmatrix}
    \phantom{-}1 & 0 & \phantom{-} 0 \\
    \phantom{-} \sin(\phi_1\alpha) \tan(\phi_2\alpha) & \phantom{-}\cos(\phi_1\alpha) & \phantom{-} \sec(\phi_2\alpha) \sin(\phi_1\alpha) \\
    \phantom{-} \cos(\phi_1\alpha) \tan(\phi_2 \alpha) & - \sin(\phi_1\alpha) & \phantom{-}\cos(\phi_1\alpha) \sec(\phi_2\alpha)
  \end{pmatrix}\,.
\end{equation}
Choosing $\alpha=0$, this background generalized vielbein describes a $S^3$ with $H$-Flux.

\subsection{\texorpdfstring{ISO($3$)}{ISO(3)}}\label{app:ISO(3)}
Applying the rotation
\begin{equation}
  R_A{}^B = \frac{1}{\sqrt{2}} \begin{pmatrix}
    \phantom{-}0 & & \phantom{-}0 & & \phantom{-}1 & & \phantom{-}0 & & \phantom{-}0 & & \phantom{-}1 \\
    \phantom{-}0 & & \phantom{-}1 & & \phantom{-}0 & & \phantom{-}0 & & \phantom{-}1 & & \phantom{-}0 \\
    -1 & & \phantom{-}0 & & \phantom{-}0 & & -1 & & \phantom{-}0 & & \phantom{-}0 \\ \phantom{-}0 & & \phantom{-}0 & & -1 & & \phantom{-}0 & & \phantom{-}0 & & \phantom{-}1 \\ \phantom{-}0 & & -1 & & \phantom{-}0 & & \phantom{-}0 & & \phantom{-}1 & & \phantom{-}0 \\
    -1 & & \phantom{-}0 & & \phantom{-}0 & & \phantom{-}1 & & \phantom{-}0 & & \phantom{-}0
  \end{pmatrix}\,,
    \quad \text{results in} \quad 
  \eta'_{AB} = \begin{pmatrix} 0 && 0 && 0 && 1 && 0 && 0 \\
    0 && 0 && 0 && 0 && 1 && 0 \\
    0 && 0 && 0 && 0 && 0 && 1 \\
    1 && 0 && 0 && 0 && 0 && 0 \\
    0 && 1 && 0 && 0 && 0 && 0 \\
    0 && 0 && 1 && 0 && 0 && 0 \\
  \end{pmatrix}
\end{equation}
and the non-semisimple Lie algebra
\begin{equation}\label{eqn:iso3generators}
  [ s_a ,s_b ] =   \cos(\alpha) \varepsilon_{ab}{}^c \, s_c + \sin(\alpha) \varepsilon_{ab}{}^c  \, t_c\,, \quad 
  [ s_a ,t_b ] =   \cos(\alpha) \varepsilon_{ab}{}^{c} \, t_{c} 
    \quad \text{and} \quad
  [ t_a ,t_b ] = 0\,,
\end{equation}
after assigning the symbols
\begin{equation}
  t_A = \{s_1, s_2, s_3, t_1, t_2, t_3\}
\end{equation}
for the generators. For $\alpha = 0$, this algebra is equivalent to $\mathfrak{iso}(3)$, which arises from a Lie algebra contraction of $\mathfrak{so}(4)$ \cite{Subag2012:contract}.

The center \eqref{eqn:center} of this algebra is trivial. Thus, the $6$-dimensional adjoint representation of the generators is faithful. In order to obtain group elements $g$, we apply the same exponential map \eqref{eqn:prodexpmap} as for SO($4$), but this time we use the coordinates
\begin{equation}
  X^I = \{\phi_1, \phi_2, \phi_3, x_1, x_2, x_3\}\,.
\end{equation}

Finally, we construct the left invariant Maurer-Cartan form for the case $\alpha=0$. We only chose this restriction to get results which are not too bulky. The procedure works for all values of $\alpha$ in the same manner. By evaluating \eqref{eqn:leftinvmc}, we obtain
\begin{equation}\label{eqn:iso3vielbein}
  E^A{}_I = \begin{pmatrix}
    A_1(\phi_1,\phi_2,\phi_3) & 0 \\
    0 & B(\phi_1,\phi_2,\phi_3)
      \end{pmatrix}
\end{equation}
and its inverse transposed
\begin{equation}
  E_A{}^I = \begin{pmatrix}
    A_1^{-T}(\phi_1,\phi_2,\phi_3) & 0 \\
    0 & B(\phi_1,\phi_2,\phi_3)
  \end{pmatrix}
\end{equation}
where $A_1(\phi_1,\phi_2,\phi_3)$ is given in \eqref{eqn:Aalpha} and $B(\phi_1,\phi_2,\phi_3)$ is defined as
\begin{equation}
  B(\phi_1,\phi_2,\phi_3) = \begin{pmatrix}
   \phantom{-}\text{c}_2 \, \text{c}_3 & \phantom{-} \text{c}_2 \, \text{s}_3 & -\text{s}_2 \\
   \phantom{-}\text{c}_3 \, \text{s}_1 \,\text{s}_2 - \text{c}_1 \, \text{s}_3 & \phantom{-}\text{c}_1 \, \text{c}_3 + \text{s}_1 \, \text{s}_2 \, \text{s}_3  & \phantom{-} \text{c}_2 \, \text{s}_1 \\
   \phantom{-} \text{c}_1 \, \text{c}_3 \, \text{s}_2 + \text{s}_1 \, \text{s}_3 & - \text{c}_3 \, \text{s}_1 + \text{c}_1 \, \text{s}_2 \, \text{s}_3  & \phantom{-}\text{c}_1 \, \text{c}_2
  \end{pmatrix} 
    \quad \text{with} \quad
  \begin{aligned}
    \text{s}_i &= \sin \phi_i \\
    \text{c}_i &= \cos \phi_i
  \end{aligned}\,.
\end{equation}

\subsection{\texorpdfstring{CSO($2,0,2$)}{CSO(2,0,2)}/\texorpdfstring{$\mathfrak{f}_1$}{f1}}\label{app:cso202}
Applying the rotation
\begin{equation}
  R_A{}^B = \frac{1}{\sqrt{2}} \begin{pmatrix}
    \phantom{-}1 & & \phantom{-}0 & & \phantom{-}0 & & \phantom{-}1 & & \phantom{-}0 & & \phantom{-}0 \\
    \phantom{-}0 & & \phantom{-}0 & & \phantom{-}0 & & \phantom{-}0 & & \phantom{-}0 & & \phantom{-}\sqrt{2} \\
    \phantom{-}0 & & \phantom{-}0 & & \phantom{-}0 & & \phantom{-}0 & & \phantom{-}\sqrt{2} & & \phantom{-}0 \\
    -1 & & \phantom{-}0 & & \phantom{-}0 & & \phantom{-}1 & & \phantom{-}0 & & \phantom{-}0 \\
    \phantom{-}0 & & \phantom{-}0 & & \phantom{-}\sqrt{2} & & \phantom{-}0 & & \phantom{-}0 & & \phantom{-}0 \\
    \phantom{-}0 & & \phantom{-}\sqrt{2} & & \phantom{-}0 & & \phantom{-}0 & & \phantom{-}0 & & \phantom{-}0
  \end{pmatrix}\,,
\end{equation}
results in
\begin{equation}
  \eta'_{AB} =  \begin{pmatrix}
    \phantom{-}0 & \phantom{-}0 & \phantom{-}0 & \phantom{-}1 & \phantom{-}0 & \phantom{-}0 \\
    \phantom{-}0 & -1 & \phantom{-}0 & \phantom{-}0 & \phantom{-}0 & \phantom{-}0 \\
    \phantom{-}0 & \phantom{-}0 & -1 & \phantom{-}0 & \phantom{-}0 & \phantom{-}0 \\
    \phantom{-}1 & \phantom{-}0 & \phantom{-}0 & \phantom{-}0 & \phantom{-}0 & \phantom{-}0 \\
    \phantom{-}0 & \phantom{-}0 & \phantom{-}0 & \phantom{-}0 & \phantom{-}1 & \phantom{-}0 \\
    \phantom{-}0 & \phantom{-}0 & \phantom{-}0 & \phantom{-}0 & \phantom{-}0 & \phantom{-}1
  \end{pmatrix}
\end{equation}
and the solvable Lie algebra
\begin{align}\label{eqn:cso202generators}
  [ t_0 ,t_a ] &= \alpha_+ \, \varepsilon_a{}^b \, t_b\,, & & &
  [ t_0 ,s_a ] &= \phantom{-} \alpha_- \, \varepsilon_a{}^b \, s_b\,, \nonumber \\
  [ t_a ,t_b ] &= \alpha_+ \, \varepsilon_{ab} \, z  & &\text{and}&
  [ s_a ,s_b ] &= -\alpha_- \, \varepsilon_{ab} \, z
\intertext{with}
  \alpha_+ &= -\cos(\alpha) - \sin(\alpha) 
    & &\text{and}&
  \alpha_- &=  \cos(\alpha) - \sin(\alpha)
\end{align}
after assigning the symbols
\begin{equation}
  t_A = \{t_0, s_1, s_2, z, t_1, t_2\}
\end{equation}
for the generators. The indices $a, b, c, \dots$ run from $1$ to $2$ and $\varepsilon_a{}^b$ denotes the totally antisymmetric tensor in two dimensions with $\varepsilon_1{}^2 = 1$. For $\alpha=0$, this algebra is equivalent to $\mathfrak{cso}(2,0,2)$. Its derived series reads
\begin{equation}
  L^0 = \{t_0, t_1, t_2, s_1, s_2, z\} \supset \{t_1, t_2, s_1, s_2, z\} \supset \{z\} \supset \{0\}
\end{equation}
for $\alpha\ne \pi/4$ and $z$ is the non-trivial center. Thus, the adjoint representation is not faithful. As explained in section~\ref{sec:solvable}, we read off the nilpotent subalgebra
\begin{equation}
  \mathfrak{n} = L^1 = \{s_1, s_2, z, t_1, t_2\} \quad \text{and the remaining generators} \quad \mathfrak{q} = \{t_0\}\,.
\end{equation}
This subalgebra gives rise to the lower central series
\begin{equation}
  L_0 = \mathfrak{n} = \{s_1, s_2, z, t_1, t_2\} \supset
  \{z\} \supset \{0\}\,,
\end{equation}
showing that $\mathfrak{n}$ is indeed nilpotent of order $k=2$.

With this data, we construct the $N=16$-dimensional subspace
\begin{align}
  V^2 = \{&s_1^2,\, s_1 s_2,\, s_2^2,\, t_1^2,\, t_1 t_2,\, s_1 t_1,\, s_2 t_1,\, t_2^2,\, s_1 t_2,\, s_2 t_2, z, &&\ord \cdot = 2 \nonumber \\
    &t_1,\, t_2,\, s_1,\, s_2, &&\ord \cdot = 1 \nonumber \\
    &1 \} &&\ord \cdot = 0
\end{align}
of the universal enveloping algebra. We obtain the generators by following the procedure outlined in section~\ref{sec:solvable}.

Group elements arise from the exponential map \eqref{eqn:prodexpmap} with the coordinates
\begin{equation}\label{eqn:cso202coord}
  X^I = \{ \phi, x_1, x_2, z, y_1, y_2 \}
\end{equation}
We calculate the background generalized vielbein by using \ref{eqn:leftinvmc} and obtain
\begin{equation}
  E^A{}_I = \begin{pmatrix}
    \phantom{-}1 & \phantom{-}0 & \phantom{-}0 & \phantom{-}0 & \phantom{-}0 & \phantom{-}0 \\
    \phantom{-}0 & \phantom{-}\cos(\alpha_- \phi) & \phantom{-}\sin(\alpha_-  \phi) & \phantom{-}0 & \phantom{-}0 & \phantom{-}0 \\
    \phantom{-}0 & -\sin(\alpha_- \phi) & \phantom{-}\cos(\alpha_- \phi) & \phantom{-}0 & \phantom{-}0 & \phantom{-}0 \\
    \phantom{-}0 & \phantom{-}0 & \alpha_- \, x_1 & \phantom{-}1 & \phantom{-}0 & -\alpha_+ \, y_1 \\ 
    \phantom{-}0 & \phantom{-}0 & \phantom{-}0 & \phantom{-}0 & \phantom{-}\cos(\alpha_+ \, \phi) & \phantom{-}\sin(\alpha_+ \, \phi) \\
    \phantom{-}0 & \phantom{-}0 & \phantom{-}0 & \phantom{-}0 & -\sin(\alpha_+ \, \phi) & \phantom{-}\cos( \alpha_+ \, \phi)
  \end{pmatrix}
\end{equation}
with the inverse transposed
\begin{equation}
E_A{}^I = \begin{pmatrix}
    \phantom{-}1 & \phantom{-}0 & \phantom{-}0 & \phantom{-}0 & \phantom{-}0 & \phantom{-}0 \\
    \phantom{-}0 & \phantom{-}\cos(\alpha_- \phi) & \phantom{-}\sin(\alpha_-  \phi) & -\alpha_- \, x_1 \, \sin(\alpha_- \phi) & \phantom{-}0 & \phantom{-}0 \\
    \phantom{-}0 & -\sin(\alpha_- \phi) & \phantom{-}\cos(\alpha_- \phi) & -\alpha_- \, x_1 \, \cos(\alpha_- \phi) & \phantom{-}0 & \phantom{-}0 \\
    \phantom{-}0 & \phantom{-}0 & \phantom{-}0 & \phantom{-}1 & \phantom{-}0 & \phantom{-}0 \\ 
    \phantom{-}0 & \phantom{-}0 & \phantom{-}0 & \phantom{-}\alpha_+ \, y_1 \, \sin(\alpha_+ \phi) & \phantom{-}\cos(\alpha_+ \, \phi) & \phantom{-}\sin(\alpha_+ \, \phi) \\
    \phantom{-}0 & \phantom{-}0 & \phantom{-}0 & \phantom{-}\alpha_+ \, y_1 \, \cos(\alpha_+ \phi) & -\sin(\alpha_+ \, \phi) & \phantom{-}\cos( \alpha_+ \, \phi)
  \end{pmatrix}\,.
\end{equation}
This background generalized vielbein describes a twisted torus. Its base is given by a circle with the coordinate $\phi$. Over this circle, a two dimensional torus is fibered. The monodromy, which arises after one complete cycle around the base, can be expressed in terms of the complex structure / K\"ahler parameter of the fibered torus. There are two important cases: First, $\alpha=0$ give rise to a geometric solve manifold. It is also call single elliptic case. Secondly, $\alpha\ne 0$ corresponds to the double elliptic case \cite{Hassler:2014sba}. This background is not T-dual to any geometric configuration. For $\alpha =\pm \pi/4$ the group reduces to $\mathfrak{f}_1$ as $\alpha_+$ or $\alpha_-$ becomes zero \cite{Dibitetto:2012rk}.

\subsection{\texorpdfstring{$\mathfrak{h}_1$}{h1}}
Applying the rotation
\begin{equation}
  R_A{}^B = \frac{1}{\sqrt{2}} \begin{pmatrix}
    -1 & & \phantom{-}0 & & \phantom{-}0 & & \phantom{-}1 & & \phantom{-}0 & & \phantom{-}0 \\
    \phantom{-}1 & & \phantom{-}0 & & \phantom{-}0 & & \phantom{-}1 & & \phantom{-}0 & & \phantom{-}0 \\
    \phantom{-}0 & & -1 & & \phantom{-}0 & & \phantom{-}0 & & \phantom{-}0 & & \phantom{-}1 \\
    \phantom{-}0 & & \phantom{-}0 & & -1 & & \phantom{-}0 & & -1 & & \phantom{-}0 \\
    \phantom{-}0 & & \phantom{-}1 & & \phantom{-}0 & & \phantom{-}0 & & \phantom{-}0 & & \phantom{-}1 \\
    \phantom{-}0 & & \phantom{-}0 & & -1 & & \phantom{-}0 & & \phantom{-}1 & & \phantom{-}0
  \end{pmatrix}\,,
    \quad \text{results in} \quad
  \eta'_{AB} = \begin{pmatrix}
    \phantom{-}0 & \phantom{-}0 & \phantom{-}0 & -1 & \phantom{-}0 & \phantom{-}0 \\
    \phantom{-}0 & \phantom{-}0 & \phantom{-}0 & \phantom{-}0 & \phantom{-}0 & \phantom{-}1 \\
    \phantom{-}0 & \phantom{-}0 & \phantom{-}0 & \phantom{-}0 & -1 & \phantom{-}0 \\ -1 & \phantom{-}0 & \phantom{-}0 & \phantom{-}0 & \phantom{-}0 & \phantom{-}0 \\
    \phantom{-}0 & \phantom{-}0 & -1 & \phantom{-}0 & \phantom{-}0 & \phantom{-}0 \\
    \phantom{-}0 & \phantom{-}1 & \phantom{-}0 & \phantom{-}0 & \phantom{-}0 & \phantom{-}0
  \end{pmatrix}
\end{equation}
and the solvable Lie algebra
\begin{align}
  [t_0, t_a] &= \phantom{-}\cos(\alpha) \, \varepsilon_a{}^b \, t_b\,, & &&
  [t_0, s_a] &= \phantom{-}\cos(\alpha) \, \varepsilon_a{}^b \, s_b - \sin(\alpha) \, t_a\,, \\
  [s_a, s_b] &= -\sin(\alpha)\, \varepsilon_{ab} \, z   & &\text{and}&
  [t_a, s_b] &=  -\cos(\alpha) \, \delta_{ab} \, z\,,
\end{align}
after assigning the symbols
\begin{equation}
  t_A = \{t_0, s_1, s_2, z, t_1, t_2\}\,.
\end{equation}
Both, its derived series and the lower central series of its nilpotent Lie subalgebra $\mathfrak{n}$, match with the $\mathfrak{cso}(2,0,2)$ case discussed in the last subsection. Thus, obtaining the $N=16$-dimensional matrix representation of the generators, goes exactly along the lines of appendix~\ref{app:cso202}. 

Group elements arise from the exponential map \eqref{eqn:prodexpmap} with the coordinates given in \eqref{eqn:cso202coord}. Here, we only present the  background generalized vielbein for $\alpha=0$. In this case, we recover the $\mathfrak{h}_1$ algebra presented in \cite{Dibitetto:2012rk}. This restriction is not mandatory, however, it simplifies the results
\begin{equation}
  E^A{}_I = \begin{pmatrix}
  \phantom{-}1 & \phantom{-}0 & \phantom{-}0 & \phantom{-}0 & \phantom{-}0 & \phantom{-}0 \\ \phantom{-}0 & \phantom{-}\cos(\phi) & \phantom{-}\sin(\phi) & \phantom{-}0 & \phantom{-}0 & \phantom{-}0 \\ \phantom{-}0 & -\sin(\phi) & \phantom{-}\cos(\phi) & \phantom{-}0 & \phantom{-}0 & \phantom{-}0 \\ \phantom{-}0 & \phantom{-}0 & \phantom{-}0 & \phantom{-}1 & -x_1 & -x_2 \\ \phantom{-}0 & \phantom{-}0 & \phantom{-}0 & \phantom{-}0 & \phantom{-}\cos(\phi) & \phantom{-}\sin(\phi) \\ \phantom{-}0 & \phantom{-}0 & \phantom{-}0 & \phantom{-}0 & -\sin(\phi) & \phantom{-}\cos(\phi) \\ 
  \end{pmatrix}
\end{equation}
and its inverse transposed
\begin{equation}
  E_A{}^I = \begin{pmatrix}
  \phantom{-}1 & \phantom{-}0 & \phantom{-}0 & \phantom{-}0 & \phantom{-}0 & \phantom{-}0 \\ \phantom{-}0 & \phantom{-}\cos(\phi) & \phantom{-}\sin(\phi) & \phantom{-}0 & \phantom{-}0 & \phantom{-}0 \\ \phantom{-}0 & -\sin(\phi) & \phantom{-}\cos(\phi) & \phantom{-}0 & \phantom{-}0 & \phantom{-}0 \\ \phantom{-}0 & \phantom{-}0 & \phantom{-}0 & \phantom{-}1 & \phantom{-}0 & \phantom{-}0 \\ \phantom{-}0 & \phantom{-}0 & \phantom{-}0 & x_1 \, \cos(\phi) + x_2 \, \sin(\phi) & \phantom{-}\cos(\phi) & \phantom{-}\sin(\phi) \\ \phantom{-}0 & \phantom{-}0 & \phantom{-}0 & x_2 \, \cos(\phi) - x_1 \, \sin(\phi) & -\sin(\phi) & \phantom{-}\cos(\phi) \\ 
  \end{pmatrix}
\end{equation}
considerably.

\subsection{\texorpdfstring{CSO($1,0,3$)}{CSO(1,0,3)}/\texorpdfstring{$\mathfrak{l}$}{l}}
Applying the rotation
\begin{equation}
  R_A{}^B = \frac{1}{\sqrt{2}} \begin{pmatrix}
    \phantom{-}0 & & -1 & & \phantom{-}0 & & \phantom{-}0 & & \phantom{-}1 & & \phantom{-}0 \\
    -1 & & \phantom{-}0 & & \phantom{-}0 & & \phantom{-}1 & & \phantom{-}0 & & \phantom{-}0 \\
    \phantom{-}0 & & \phantom{-}0 & & \cos(\alpha) + \sin(\alpha) & & \phantom{-}0 & & \phantom{-}0 & & -\cos(\alpha) + \sin(\alpha)
\\
    \phantom{-}0 & & \phantom{-}1 & & \phantom{-}0 & & \phantom{-}0 & & \phantom{-}1 & & \phantom{-}0 \\ \phantom{-}1 & & \phantom{-}0 & & \phantom{-}0 & & \phantom{-}1 & & \phantom{-}0 & & \phantom{-}0 \\
    \phantom{-}0 & & \phantom{-}0 & & -\cos(\alpha) + \sin(\alpha) & & \phantom{-}0 & & \phantom{-}0 & & -\cos(\alpha) - \sin(\alpha)
  \end{pmatrix}\,,
\end{equation}
results in
\begin{equation}\label{eqn:etacso(103)}
  \eta'_{AB} = \begin{pmatrix}
    \phantom{-}0 & \phantom{-}0 & \phantom{-}0 & -1 & \phantom{-}0 & \phantom{-}0 \\
    \phantom{-}0 & \phantom{-}0 & \phantom{-}0 & \phantom{-}0 & -1 & \phantom{-}0 \\
    \phantom{-}0 & \phantom{-}0 & \phantom{-}\sin(2\alpha) & \phantom{-}0 & \phantom{-}0 & -\cos(2\alpha) \\
    -1 & \phantom{-}0 & \phantom{-}0 & \phantom{-}0 & \phantom{-}0 & \phantom{-}0 \\
    \phantom{-}0 & -1 & \phantom{-}0 & \phantom{-}0 & \phantom{-}0 & \phantom{-}0 \\
    \phantom{-}0 & \phantom{-}0 & -\cos(2\alpha) & \phantom{-}0 & \phantom{-}0 & -\sin(2\alpha)
  \end{pmatrix}
\end{equation}
and the nilpotent Lie algebra
\begin{equation}
\label{eqn:cso103commutator}
  [t_1, t_2 ] = \cos(2\alpha) \, z_3 - \sin(2\alpha) \, t_3\,, \quad [t_2, t_3] = z_1 \quad \text{and} \quad [t_3, t_1] =  z_2
\end{equation}
after assigning the symbols
\begin{equation}
  t_A = \{t_1, t_2, t_3, z_1, z_2, z_3\}\,.
\end{equation}
For $\alpha=0$, we obtain the Lie algebra
\begin{equation}
  [t_a, t_b] = \varepsilon_{ab}{}^c \, z_c
\end{equation}
which is called $\mathfrak{cso}(1,0,3)$ \cite{Dibitetto:2012rk}. It is nilpotent of order 2 and its lower central series reads
\begin{equation}
  L_0 = \{ t_1, t_2, t_3, z_1, z_2, z_3 \} \supset \{ z_1, z_2, z_3 \} \supset \{0\}\,.
\end{equation}
The center of this algebra is $\{ z_1, z_2, z_3 \}$. Following the procedure outlined in section~\ref{sec:nilpotent}, we construct the $N=13$-dimensional subspace
\begin{align}
  V^2 = \{&t_1^2,\, t_1 t_2,\, t_1 t_3,\, t_2^2,\, t_2 t_3,\, t_3^2,\, z^1,\, z^2,\, z^3, &&\ord \cdot = 2 \nonumber \\
    &t_1,\, t_2,\, t_3, &&\ord \cdot = 1 \nonumber \\
    &1 \} &&\ord \cdot = 0
\end{align}
of the universal enveloping algebra. Finally, we obtain the matrix representation of the generators $t_A$, by expanding the linear maps $\phi_{t_A}$ in the basis spanned by $V^2$. Group elements are derived from the exponential map \eqref{eqn:prodexpmap} using the coordinates
\begin{equation}
  X^I = \{ x_1, x_2, x_3, z_1, z_2, z_3 \}\,.
\end{equation}
They give rise to the left-invariant Maurer-Cartan form
\begin{equation}
  E^A{}_I = \begin{pmatrix}
    \phantom{-}1 && \phantom{-}0 && \phantom{-}0 && \phantom{-}0 && \phantom{-}0 && \phantom{-}0 \\
    \phantom{-}0 && \phantom{-}1 && \phantom{-}0 && \phantom{-}0 && \phantom{-}0 && \phantom{-}0 \\
    \phantom{-}0 && \phantom{-}0 && \phantom{-}1 && \phantom{-}0 && \phantom{-}0 && \phantom{-}0 \\
    \phantom{-}0 && \phantom{-}0 && -x_2 && \phantom{-}1 && \phantom{-}0 && \phantom{-}0 \\ \phantom{-}0 && \phantom{-}0 && \phantom{-}x_1 && \phantom{-}0 && \phantom{-}1 && \phantom{-}0 \\
    \phantom{-}0 && -x_1 && \phantom{-}0 && \phantom{-}0 && \phantom{-}0 && \phantom{-}1 \\
  \end{pmatrix}
\end{equation}
with the inverse transposed
\begin{equation}
  E_A{}^I = \begin{pmatrix}
    \phantom{-}1 && \phantom{-}0 && \phantom{-}0 && \phantom{-}0 && \phantom{-}0 && \phantom{-}0 \\
    \phantom{-}0 && \phantom{-}1 && \phantom{-}0 && \phantom{-}0 && \phantom{-}0 && \phantom{-}x_1 \\
    \phantom{-}0 && \phantom{-}0 && \phantom{-}1 && \phantom{-}x_2 && -x_1 && \phantom{-}0 \\
    \phantom{-}0 && \phantom{-}0 && \phantom{-}0 && \phantom{-}1 && \phantom{-}0 && \phantom{-}0 \\ \phantom{-}0 && \phantom{-}0 && \phantom{-}0 && \phantom{-}0 && \phantom{-}1 && \phantom{-}0 \\
    \phantom{-}0 && \phantom{-}0 && \phantom{-}0 && \phantom{-}0 && \phantom{-}0 && \phantom{-}1 \\
  \end{pmatrix}\,.
\end{equation}
This background generalized vielbein describes a $3$-torus with $H$-flux.

For $\alpha\ne0$, the lower central series changes
\begin{equation}
  L_0 = \{ t_1, t_2, t_3', z_1, z_2, z_3' \} \supset \{ z_1, z_2, z_3' \} \supset \{z_1, z_2\} \supset \{0\}\,,
\end{equation}
where we have introduced the abbreviations
\begin{equation}\label{eqn:t3z3bar}
  t_3' = \cos(2\alpha) t_3 - \sin(2\alpha) z_3 \quad \text{and} \quad
  z_3' = \sin(2\alpha) t_3 + \cos(2\alpha) z_3\,,
\end{equation}
identifying a nilpotent Lie algebra of order $3$. If we want to treat it in a proper way, we have to extend $V^2$ to
\begin{align}
  V^3 = \{& t_1^3,\, t_1^2 t_2,\, t_1^2 t_3',\,t_1 t_2^2,\, t_1 t_2 t_3,\,
    t_1 {t_3}'^2,\, t_2 {t_3}'^2,\, {t_3}'^3,\, t_1 z_3',\, t_2 z_3',\,   t_3' z_3',\, z_1,\, z_2, &&\ord \cdot = 3 \nonumber \\  
    & t_1^2,\, t_1 t_2,\, t_1 t_3',\, t_2^2,\, t_2 t_3',\, {t_3'}^2,\, z_3', &&\ord \cdot = 2 \nonumber \\
    &t_1,\, t_2,\, t_3', &&\ord \cdot = 1 \nonumber \\
    &1 \}\,. &&\ord \cdot = 0
\end{align} They give rise to the modified Lie algebra
\begin{align}
  [t_1, t_2]  &= z_3'\,, & [t_1, z_3'] &= \sin(2 \alpha) z_2\,, & [z_3', t_2] &= \sin(2 \alpha) z_1\,, \\
  [t_2, t_3'] &= \cos(2 \alpha) z_1 & & \text{and} & [t_3', t_1] &= \cos(2 \alpha) z_2\,,
\end{align}
which we have to use to evaluate the map $\phi_{t_A}$ in the basis $V^3$. Doing so, we obtain a $N=24$-dimensional matrix representation for the generators of the Lie algebra. Exponentiating them according to \eqref{eqn:prodexpmap} and using the arising group elements to calculate the  background generalized vielbein, one obtains
\begin{equation}
  E^A{}_I = \begin{pmatrix} \phantom{-}1 & \phantom{-}0 & \phantom{-}0 & \phantom{-}0 & \phantom{-}0 & \phantom{-}0 \\
  \phantom{-}0 & \phantom{-}1 & \phantom{-}0 & \phantom{-}0 & \phantom{-}0 & \phantom{-}0  \\ \phantom{-}0 & \phantom{-}0 & \phantom{-}1 & \phantom{-}0 & \phantom{-}0 & \phantom{-}0  \\ \phantom{-}0 & \phantom{-}0 & -x_2 \, \cos(2\alpha) & \phantom{-}1 & \phantom{-}0 & \phantom{-}x_2 \, \sin(2\alpha)  \\ \phantom{-}0 & \phantom{-}x_1^2 \, \cos(\alpha) \, \sin(\alpha)  & \phantom{-}x_1 \, \cos(2\alpha) & \phantom{-}0 & \phantom{-}1 & -x_1 \, \sin(2\alpha)  \\ \phantom{-}0 & -x_1 & \phantom{-}0 & \phantom{-}0 & \phantom{-}0 & \phantom{-}1  \end{pmatrix}
\end{equation}
assuming the coordinates
\begin{equation}
  X^I = \{ x_1, x_2, x_3', z_1, z_2, z_3' \}\,.
\end{equation}
The inverse transposed reads
\begin{equation}
E_A{}^I =\begin{pmatrix} \phantom{-}1 & \phantom{-}0 & \phantom{-}0 & \phantom{-}0 & \phantom{-}0 & \phantom{-}0 \\
  \phantom{-}0 & \phantom{-}1 & \phantom{-}0 & -x_1 \, x_2 \, \sin(2\alpha) & \phantom{-}x_1^2 \, \cos(\alpha) \, \sin(\alpha) & \phantom{-}x_1  \\ \phantom{-}0 & \phantom{-}0 & \phantom{-}1 & \phantom{-}x_2 \, \cos(2\alpha) & -x_1 \, \cos(2\alpha) & \phantom{-}0  \\ \phantom{-}0 & \phantom{-}0 & \phantom{-}0 & \phantom{-}1 & \phantom{-}0 & \phantom{-}0  \\ \phantom{-}0 & \phantom{-}0  & \phantom{-}0 & \phantom{-}0 & \phantom{-}1 & \phantom{-}0  \\ \phantom{-}0 & \phantom{-}0 & \phantom{-}0 & -x_2 \, \sin(2\alpha) & \phantom{-}x_1 \, \sin(2\alpha) & \phantom{-}1  \end{pmatrix}\,.
\end{equation}
Flat indices, such as the $A$ of $E^A{}_I$, are lowered with the $\eta_{AB}$ metric 
\begin{equation}
  \eta''_{AB} = \begin{pmatrix}
    \phantom{-}0 & \phantom{-}0 & \phantom{-}0 & -1 & \phantom{-}0 & \phantom{-}0 \\
    \phantom{-}0 & \phantom{-}0 & \phantom{-}0 & \phantom{-}0 & -1 & \phantom{-}0 \\
    \phantom{-}0 & \phantom{-}0 & -\sin(2\alpha) & \phantom{-}0 & \phantom{-}0 & -\cos(2\alpha) \\
    -1 & \phantom{-}0 & \phantom{-}0 & \phantom{-}0 & \phantom{-}0 & \phantom{-}0 \\
    \phantom{-}0 & -1 & \phantom{-}0 & \phantom{-}0 & \phantom{-}0 & \phantom{-}0 \\
    \phantom{-}0 & \phantom{-}0 & -\cos(2\alpha) & \phantom{-}0 & \phantom{-}0 & \phantom{-}\sin(2\alpha)
  \end{pmatrix}\,.
\end{equation}
For $\alpha=\pi/4$, we find the algebra $\mathfrak{l}$ presented in \cite{Dibitetto:2012rk}, after an additional rotation of the $\mathfrak{cso}(1,0,3)$ structure coefficients \eqref{eqn:cso103commutator} with
\begin{equation}
  R''_A{}^B = \begin{pmatrix}
    -1 & \phantom{-}0 & \phantom{-}0 & \phantom{-}0 & \phantom{-}0 & \phantom{-}0 \\
    \phantom{-}0 & \phantom{-}1 & \phantom{-}0 & \phantom{-}0 & \phantom{-}0 & \phantom{-}0 \\
    \phantom{-}0 & \phantom{-}0 & \phantom{-}0 & \phantom{-}1 & \phantom{-}0 & \phantom{-}0 \\
    \phantom{-}0 & \phantom{-}0 & \phantom{-}0 & \phantom{-}0 & \phantom{-}0 & \phantom{-}1 \\
    \phantom{-}0 & \phantom{-}0 & \phantom{-}0 & \phantom{-}0 & \phantom{-}1 & \phantom{-}0 \\
    \phantom{-}0 & \phantom{-}0 & \phantom{-}1 & \phantom{-}0 & \phantom{-}0 & \phantom{-}0
  \end{pmatrix}
    \quad \text{resulting in} \quad 
  \eta'''_{AB} = \begin{pmatrix}
    \phantom{-}0 & \phantom{-}0 & \phantom{-}0 & \phantom{-}0 & \phantom{-}0 & \phantom{-}1  \\
    \phantom{-}0 & \phantom{-}0 & \phantom{-}0 & \phantom{-}0 & -1 & \phantom{-}0  \\
    \phantom{-}0 & \phantom{-}0 & -1 & \phantom{-}0 & \phantom{-}0 & \phantom{-}0  \\
    \phantom{-}0 & \phantom{-}0 & \phantom{-}0 & \phantom{-}1 & \phantom{-}0 & \phantom{-}0 \\
    \phantom{-}0 & -1 & \phantom{-}0 & \phantom{-}0 & \phantom{-}0 & \phantom{-}0  \\
    \phantom{-}1 & \phantom{-}0 & \phantom{-}0 & \phantom{-}0 & \phantom{-}0 & \phantom{-}0 
  \end{pmatrix}
\end{equation}
and the commutator relations
\begin{equation}
  \big[t_1, t_2 \big] = t_4 \quad \big[t_1, t_4 \big] =  t_5 \quad \big[t_2, t_4 \big] =  t_6
\end{equation}
where we assigned the symbols
\begin{equation}
  t_A = \{ t_1, t_2, t_3, t_4, t_5, t_6 \}
\end{equation}
for the generators.

\bibliography{literatur}
\bibliographystyle{JHEP}
\end{document}